\pdfoutput=1
\documentclass{emulateapj}
\newcommand{\etal}{et~al.}
\usepackage{color}

\newcommand{\Lsun}{L$_{\odot}$}
\newcommand{\Msun}{M$_{\odot}$}

\newcommand{\cgsflux}{erg~s$^{-1}$~cm$^{-2}$}
\newcommand{\cgsflam}{erg~s$^{-1}$~cm$^{-2}$~\AA$^{-1}$}

\newcommand{\kms}{\hbox{km~s$^{-1}$}}

\newcommand{\ArII}{\hbox{{\rm Ar}\kern 0.1em{\sc ii}}}
\newcommand{\ArIII}{\hbox{{\rm Ar}\kern 0.1em{\sc iii}}}
\newcommand{\CIV}{\hbox{{\rm C}\kern 0.1em{\sc iv}}}
\newcommand{\HI}{\hbox{{\rm H}\kern 0.1em{\sc i}}}
\newcommand{\HII}{\hbox{{\rm H}\kern 0.1em{\sc ii}}}
\newcommand{\HeI}{\hbox{{\rm He}\kern 0.1em{\sc i}}}
\newcommand{\HeII}{\hbox{{\rm He}\kern 0.1em{\sc ii}}}
\newcommand{\NII}{\hbox{{\rm N}\kern 0.1em{\sc ii}}}
\newcommand{\OI}{\hbox{{\rm O}\kern 0.1em{\sc i}}}
\newcommand{\OII}{\hbox{{\rm O}\kern 0.1em{\sc ii}}}
\newcommand{\OIII}{\hbox{{\rm O}\kern 0.1em{\sc iii}}}
\newcommand{\OIIlong}{{\rm O}\kern 0.1em{\sc ii}~$\lambda 3727$} 
\newcommand{\FeII}{\hbox{{\rm Fe}\kern 0.1em{\sc ii}}}
\newcommand{\NeII}{\hbox{{\rm Ne}\kern 0.1em{\sc ii}}}
\newcommand{\NeIII}{\hbox{{\rm Ne}\kern 0.1em{\sc iii}}}
\newcommand{\NeV}{\hbox{{\rm Ne}\kern 0.1em{\sc v}}}
\newcommand{\SII}{\hbox{{\rm S}\kern 0.1em{\sc ii}}}
\newcommand{\SIII}{\hbox{{\rm S}\kern 0.1em{\sc iii}}}
\newcommand{\SIV}{\hbox{{\rm S}\kern 0.1em{\sc iv}}}
\newcommand{\SiIV}{\hbox{{\rm Si}\kern 0.1em{\sc iv}}}
\newcommand{\MgII}{\hbox{{\rm Mg}\kern 0.1em{\sc ii}}}
\newcommand{\Halpha}{\hbox{{\rm H}\kern 0.1em$\alpha$}}
\newcommand{\Hbeta}{\hbox{{\rm H}\kern 0.1em$\beta$}}
\newcommand{\Heopta}{\hbox{{\rm He}\kern 0.1em{\sc i}}~$6678$}
\newcommand{\Heoptb}{\hbox{{\rm He}\kern 0.1em{\sc i}}~$5876$}
\newcommand{\Heoptc}{\hbox{{\rm He}\kern 0.1em{\sc i}}~$4471$}
\newcommand{\Brgam}{\hbox{{\rm Br}\kern 0.1em$\gamma$}}
\newcommand{\Brten}{\hbox{{\rm Br}\kern 0.1em$10$}}
\newcommand{\Breleven}{\hbox{{\rm Br}\kern 0.1em$11$}}
\newcommand{\HeIh}{\hbox{{\rm He}\kern 0.1em{\sc i}}~$1.7$~{\micron}}
\newcommand{\HeIk}{\hbox{{\rm He}\kern 0.1em{\sc i}}~$2.06$~{\micron}}
\newcommand{\squishlist}{
   \begin{list}{$\bullet$}
    { \setlength{\itemsep}{0pt}      \setlength{\parsep}{1pt}
      \setlength{\topsep}{3pt}       \setlength{\partopsep}{0pt}
      \setlength{\leftmargin}{1.5em} \setlength{\labelwidth}{1em}
      \setlength{\labelsep}{0.5em} } }
\newcommand{\squishend}{
    \end{list}  }

\newcommand{\clustername}{SDSSJ~1438+1454}
\newcommand{\arcname}{SGAS~1438}
\newcommand{\arcnamelong}{SGAS~143845.1$+$145407}

\newcommand{\magN}{$6.5 ^{+1.8}_{-2.5}$}
\newcommand{\magS}{$5.2 ^{+1.6}_{-2.1}$}
\newcommand{\magW}{$4.0 ^{+2.5}_{-1.3}$}
\newcommand{\magE}{$2.9 ^{+1.4}_{-0.8}$}
\newcommand{\magall}{$16.8 ^{+5.2}_{-6.2}$}
\newcommand{\magcoreN}{$4.6 ^{+1.7}_{-1.8}$}
\newcommand{\magcoreS}{$5.8 ^{+2.1}_{-2.2}$}

\newcommand{\Px}{$7.7_{-4.9}^{+2.3}$}
\newcommand{\Py}{$-2.2_{-0.8}^{+1.6}$}
\newcommand{\Pe}{$0.54_{-0.30}^{+0.06}$}
\newcommand{\Ptheta}{$165_{-  4}^{+  4}$}
\newcommand{\Prc}{$ 12_{-  6}^{+  1}$}
\newcommand{\Psigma}{$672_{-222}^{+ 50}$}
\newcommand{\Prcb}{$1.3_{-1.0}^{+2.0}$}
\newcommand{\Psigmab}{$267_{- 30}^{+ 71}$}
\newcommand{\Pcutb}{$ 57_{- 46}^{+ 23}$}

\slugcomment{}
\shorttitle{A Big Cool Starburst at $z=0.816$}
\shortauthors{Gladders~\etal}

\begin{document}

\title{\arcnamelong : A Big, Cool Starburst at Redshift $0.816$\footnote{This paper includes data gathered with the 6.5 meter Magellan Telescopes located at Las Campanas Observatory, Chile. This paper is also based on observations made with the Nordic Optical Telescope, operated on the island of La Palma jointly by Denmark, Finland, Iceland, Norway, and Sweden, in the Spanish Observatorio del Roque de los Muchachos of the Instituto de Astrof\'{i}sica de Canarias.
}}

\author{
Michael~D.~Gladders\altaffilmark{1,2}, Jane~R.~Rigby\altaffilmark{3}, Keren~Sharon\altaffilmark{2}, Eva~Wuyts\altaffilmark{1,2}, Louis~E.~Abramson\altaffilmark{1,2}, H\aa kon~Dahle\altaffilmark{4}, S.~E.~Persson\altaffilmark{5}, Andrew~J.~Monson\altaffilmark{5}, Daniel~D.~Kelson\altaffilmark{5}, Dominic~J.~Benford\altaffilmark{3}, David~Murphy\altaffilmark{5}, Matthew~B.~Bayliss\altaffilmark{6,7}, Keely~D.~Finkelstein\altaffilmark{8}, Benjamin~P.~Koester\altaffilmark{9}, Alissa~Bans\altaffilmark{1}, Eric~J.~Baxter\altaffilmark{1}, Jennifer~E.~Helsby\altaffilmark{1,2}
}

\altaffiltext{1}{Department of Astronomy \& Astrophysics, The University of Chicago, 5640 S.~Ellis Avenue, Chicago, IL 60637, USA}
\altaffiltext{2}{Kavli Institute for Cosmological Physics at the University of Chicago}
\altaffiltext{3}{Observational Cosmology Lab, NASA Goddard Space Flight Center, Greenbelt MD 20771, USA}
\altaffiltext{4}{Institute of Theoretical Astrophysics, University of Oslo, P.O. Box 1029, Blindern, N-0315, Oslo, Norway}
\altaffiltext{5}{Carnegie Observatories, 813 Santa Barbara St., Pasadena, CA 91101, USA}
\altaffiltext{6}{Harvard-Smithsonian Center for Astrophysics, 60 Garden Street, Cambridge, MA 02138, USA}
\altaffiltext{7}{Department of Physics, Harvard University, 17 Oxford St, Cambridge, MA
02138}
\altaffiltext{8}{Department of Astronomy, University of Texas at Austin, 2515 Speedway Stop C1400, Austin, TX 78712, USA}
\altaffiltext{9}{Physics Department, University of Michigan, Ann Arbor, MI 48109, USA}

\email{gladders@oddjob.uchicago.edu}

\begin{abstract}
We present the discovery and a detailed multi-wavelength study of a strongly-lensed luminous infrared galaxy at z=0.816. Unlike most known lensed galaxies discovered at optical or near-infrared wavelengths this lensed source is red, $(r-K_s)_{AB}=3.9$, which the data presented here demonstrate is due to ongoing dusty star formation. The overall lensing magnification (a factor of 17) facilitates observations from the blue optical through to 500~\micron, fully capturing both the stellar photospheric emission as well as the re-processed thermal dust emission. We also present optical and near-IR spectroscopy.  These extensive data show that this lensed galaxy is in many ways typical of IR-detected sources at $z\sim1$, with both a total luminosity and size in accordance with other (albeit much less detailed) measurements in samples of galaxies observed in deep fields with the Spitzer telescope.   Its far-infrared spectral energy distribution is well--fit by local templates that are an order of magnitude less luminous than the lensed galaxy; local templates of comparable luminosity are too hot to fit.  Its size ($D\sim7$~kpc) is much larger than local luminous infrared galaxies, but in line with sizes observed for such galaxies at $z\sim1$.  The star formation appears uniform across this spatial scale.  In this source, the luminosity of which is typical of sources that dominate the cosmic infrared background, we find that star formation is spatially extended and well organised, quite unlike the compact merger-driven starbursts which are typical for sources of this luminosity at $z\sim0$.
\end{abstract}

\keywords{galaxies: high-redshift---galaxies: evolution---gravitational lensing}

\section{Introduction}          \label{sec:intro}

Strongly gravitationally lensed galaxies offer a unique view of the distant Universe, by providing both a surfeit of photons and access to physical scales not resolvable when studying galaxies that have not been lensed.  Furthermore, lensed galaxies are likely to probe a more typical portion of the luminosity function (e.g.,\ \citealt{matt2010}) than strictly magnitude-limited blank field studies. The typical bright lensed source, from samples selected at optical wavelengths, is a vigorously star-forming galaxy at z$\sim$2 \citep{matt2011,matt2012} with blue colors and modest reddening \citep{eva2012}.   The brightest of these lensed sources, which are typically magnified by factors of 20 or more \citep[e.g.,][]{wil96,sah07,ker12}, are of particular interest in part because they enable studies in wavelength regimes where current facilities lack the sensitivity to reach high redshift galaxies (see the discussion in Siana et al. 2009).  The brightest lensed sources reported to date from discoveries at optical wavelengths are fairly blue, from the prototype MS1512-cB58  \citep{yee96} to more recent discoveries \citep[e.g.,][]{belokurov07,lin09,koester10}.  The reddest of these is the ‘Cosmic Eye’ at  $z=3.074$ \citep{smail07,siana09}.

In the submillimeter, gravitational lensing in the weak limit has provided a sensitivity boost for sub-mm galaxy surveys \citep[e.g.,][]{smail97,egami10}, but typically these sources are magnified only by tens of percent, with no multiple imaging.  As such, there were until recently only a small number of robustly confirmed gravitationally--lensed sub-mm galaxies with large magnifications \citep{borys04,kneib04,knudsen09,swinbank10,lestrada11,frayer11,combes12,fu12}. Expectations are that the new generation of millimeter and submillimeter telescopes should detect large numbers of lensed galaxies at these wavelengths (e.g., Herschel: \citealt{neg10,gon12};  the South Pole Telescope: \citealt{vieira2010}) and samples of lensed sources are now emerging from those efforts (e.g., \citealt{gre12,har12,war12}).

In this paper, we report the discovery of an unusually red and bright lensed galaxy, at a redshift of $z=0.816$. This galaxy is multiply imaged and magnified by a total factor of \magall\ by a foreground galaxy cluster at $z=0.237$. We present photometric measurements of the source from the restframe ultraviolet to the far infrared, as well as both optical and near-infrared spectroscopy. From these data we construct a lens model, which we use to invert measurements to unlensed quantities in the source plane when required.  These data show that the source is a luminous infrared galaxy (a LIRG) that is typical of star-forming galaxies at that epoch \citep{lag05,lefloch05,cap07,pas09}. \S2\ describes how the lensed source was discovered as part of a larger effort to find and characterize lensed galaxies. \S3\ describes the follow-up observations, as well as the basic measurements of the source. \S4\ details the lens model and properties of the lensed source in the image plane; these measurements are discussed and contexualised in \S5.

Throughout the paper we adopt a flat cosmology with $\Omega_M = 0.3$ and H$_0 = 70$\,km\,s$^{-1}$\,Mpc$^{-1}$. All magnitudes are reported on the AB system unless otherwise noted. 
All on-sky images show North up and East to the right.

\section{Discovery of a red herring and a red lensed galaxy} \label{sec:discovery}

The cluster \clustername\ was selected as a possible strong lensing system (M. D. Gladders et al., in preparation) from SDSS imaging. As described in Appendix A, this initial selection was likely due to the presence of a blue arclet feature 18$\farcs$5 from the brightest cluster galaxy (BCG), with color, geometry with respect to the BCG, and a hint of curvature all suggestive of lensing, as shown in Figure 1.  Appendix A also includes follow-up spectroscopy from the ARC 3.5m that shows that this initial arc candidate is not lensed, but is rather a cluster galaxy at $z=0.231$.

However, another arc-like feature, this one faint and redder than the BCG, was just visible in the SDSS imaging. We had obtained warm Spitzer observations with IRAC at 3.6 and 4.5~$\mu$m for a larger sample of $\sim100$ lensing systems through Spitzer program 70154 (PI: M.~D, Gladders). This Spitzer imaging (Figure 2, and Table 1) clearly showed that this source was infrared-bright, very red in the $g - 3.6$~\micron\ and $g - 4.5$~\micron\ colors, and likely to be a multiply imaged lensed galaxy.  This source is the subject of this paper.  Following the naming conventions established in \cite{koester10} we refer to the lensing cluster as \clustername\ and the lensed source as \arcnamelong\ (\arcname\ in short hereafter). The source name coordinates are derived from the position of the most isolated complete image, which is 5.6\arcsec\ almost due south of the BCG.

\section{Further Data and Measurements} \label{sec:data}

\subsection{Optical and Near-IR Imaging} \label{sec:oir_im}
Motivated by the identification of \arcnamelong\ as a possible lensed source from the SDSS and Spitzer imaging, we observed the field with other facilities.  Near-IR imaging in the YJH and K$_s$ bands was acquired during commissioning of the FourStar camera at the Magellan Baade 6.5m in early 2011.  These near-IR images robustly confirmed that this was indeed a lensing geometry, as shown in Figure 3, and further demonstrated that the lensed source is extremely red.   The FourStar JH and K$_s$ images were flux-calibrated directly to the Two Micron All Sky Survey point source catalog \citep{2mass}, which is possible because the field of view of FourStar is large; calibration uncertainties are a few percent in these bands. The Y-band image was calibrated using a standard star observation\footnote{Calibrations of the Y-band and other medium-band filters available in FourStar have been undertaken in support of the Z-FOURGE project \cite{spi11} and will be published in detail elsewhere.} acquired immediately prior to the science observation at a similar airmass. This calibration agrees with repeated Y-band calibrations of the instrument over months of observing to within 0.015 magnitudes; we thus consider that the Y-band data have calibration uncertainties of a few percent. In all four bands these uncertainties are sub-dominant to other sources of error. 

Optical imaging in $griz$ was obtained with the GMOS imaging spectrograph on the Gemini North 8.2m telecope; zeropoints and color terms were established by comparison to SDSS-measured stars in the same field and are uncertain by a few hundredths of a magnitude.  

The GMOS/Gemini-N and FourStar/Magellan imagery is shown in Figure 3, with images centered on the brightest cluster galaxy.  Table 1 summarizes the details of the observations and data. Photometry of the lensed source is reported in Table 1, with details of the measurement process including treatment of the central galaxy given in \S3.4.

\subsection{WISE photometry} \label{sec:wise}
\arcname\ was included in the Wide-Field Infrared Survey Explorer (WISE) preliminary release \citep{wri10}. As it is extended in the WISE bands, we perform aperture photometry rather than take the profile--fit photometry from the WISE preliminary release source catalog. Though the source is detected in the bluest two bands of WISE, these data are completely superseded by the higher spatial resolution and deeper imaging from Spitzer-IRAC, so we report here only the WISE measurements at 12~\micron\ and 22~\micron. These wavelengths are uniquely covered by WISE for this source, and are important since they measure the strength of mid-IR aromatic emission features. The details of this data and measured photometry are reported in Table 1; details of corrections for contamination by light from the BCG are given in \S3.4.

\subsection{Herschel photometry} \label{sec:herschel}
We obtained mid- and far-IR photometry for \arcname\ in discretionary time on Herschel using the PACS and SPIRE instruments, with a total clock time of 2.1 hr. Details of the observations are given in Table 1.

For PACS, we performed aperture photometry on the level 2 pipleline-produced science images, using annular apertures with radii of 10,12,16 pixels at 70~\micron, 12,14,16 pixels at 100~\micron, and 10,12,16 pixels at 160~\micron, respectively.  We applied aperture corrections as given in technical memo PICC-ME-TN-037 v1.0.  We applied color corrections as given in technical memo PICC-ME-TN-038 v1.0, by interpolating to a 27~K blackbody from the blackbodies tabulated at 20 and 30~K.\footnote{The PACS photometry is quite sensitive to the choice of color correction; we determined by fitting the photometry that a $T=27$~K blackbody was an adequate fit to the raw (not de-redshifted) photometry.   We experimented with modified blackbodies, where $F_{\nu}^{modified} = \nu^{\beta} B_{\nu}(T)$, but found these did not provide a better fit.}
These color corrections were 1.09, 1.00, and 0.97 at 70, 100, and 160~\micron.  At 70~\micron\ where the spectral slope is particularly steep, the uncertainty on the color correction is $\sim10\%$, and thus contributes significantly to the photometric uncertainty.

For SPIRE, we performed aperture photometry on the level 2 science images, using annular apertures, following the procedure of the SPIRE Photometry Cookbook.  In the Rayleigh--Jeans limit of $F_{\nu} \propto \nu^2$, the color corrections tabulated in the Cookbook are 0.94, 0.95, 0.94 at 250, 350, and 500~\micron.
Resulting PACS and SPIRE AB magnitudes are listed in Table 1, converted from measured flux densities $f_\nu$ to m$_{AB}$ = 8.9 - 2.5$\log{(f_\nu)}$, with $f_\nu$ in Janskys following the definition of AB magnitudes \citep{oke74}. 

\arcname\ is sufficiently bright in PACS and SPIRE that the photometric uncertainty is dominated not by photon noise, but by a number of systematic effects:  uncertainty in the color corection, aperture correction, and centroiding, and the fact that the source is somewhat resolved in the shortest--wavelength bands.  We estimate these effects as contributing a total of $13\%$ uncertainty on the flux density in each band.

At its band with highest spatial resolution, 70~\micron, Herschel partially resolves \arcname, in that it splits the southern image from the blended other images.  Fitting with two GALFIT components (\cite{peng10}; see \S~\ref{sec:bye_BCG}), we find that the southern image contains $22 \pm 5 \%$ of the 70~\micron\ light.  

\subsection{Subtracting the Central Galaxy and Final Photometry} \label{sec:bye_BCG}

The field of \clustername\ is complex:  the multiple images of the lensed source encircle the brightest cluster galaxy, which contributes significantly to the flux in the optical, near-infrared, and Spitzer 3.6 and 4.5~\micron\ bands, and possibly the  WISE 12~$\mu$m and 22~$\mu$m bands as well.  This complexity requires an atypically involved analysis to isolate the flux contribution of the source at each wavelength.  To do this, we fit the BCG in each of the optical and near-IR images using GALFIT \citep{peng10}. The appropriate point spread function (PSF) model for each image was constructed from fitting point sources in that image, with a set of PSF objects fit using a multi-component moffat+gaussian model in GALFIT, and then scaled and summed at a fiducial center to produce the final PSF model for each image. For the BCG fits, it was found that a multi-component S\'{e}rsic model -  with a small central component that is rounder and about 3-4 magnitudes fainter than the dominant component - was needed to achieve an acceptable fit. That the BCG requires more than one component for an acceptable fit is neither uncommon nor unexpected, and we have confirmed that this result is not due to a misconstruction of the PSF reference. The lensed source was masked iteratively, with components revealed after an initial BCG subtraction remasked and the fit iterated across all bands with the same mask. Figure 4 shows the final BCG-subtracted images for a few representative wavelengths.

The resulting BCG models do not show any significant evidence for color gradients in the BCG, and so to subtract the BCG flux from the Spitzer 3.6~$\mu$m and 4.5~$\mu$m images we took the best fit K$_s$-band BCG model, convolved it with a PSF constructed as above for the IRAC images, and then subtracted a scaled version of the resulting BCG image by eye. We did not use GALFIT to perform the subtraction automatically because the Spitzer-IRAC PSF badly blends the lens with some of the images of the source and thus the minimization calculation in GALFIT performs poorly absent more extensive masking of the lensed source.

For the IRAC data, and all images at shorter wavelengths, we then measured the magnitude of the isolated southern image of the source using the techniques described in \cite{eva2010}. We also measured the ratio of the total flux of the source to this southern image in the four reddest bands, where the flux ratio of \arcname\ to the BCG is highest.  
This ratio is 3.49$\pm$0.17. Final magnitudes reported in Table 1 are constructed from the measured magnitudes of the isolated southern image rescaled to the total image using this ratio. The simpler approach of reporting total source magnitudes in the BCG-subtracted images was avoided due to the possible influence of systematics from the BCG model subtraction. At bluer wavelengths the BCG is 4-5 magnitudes brighter than the source, such that aperture measurements of the entire source would be completely dominated by the BCG; by scaling from measurements of the isolated southern image, we trade these potential systematics which are difficult to quantify for a noisier measurement and a quantified systematic uncertainty in the scaling of less than 5\%. Note that we do not include this systematic 5\% uncertainty in Table 1; it is only relevant when comparing results at shorter wavelengths to data from Herschel and WISE, and since the spectral energy distribution (SED) fitting (see \S4.4) is done independently for the shorter and longer wavelengths it has no effect on our conclusions.

At the long wavelengths sampled by Herschel, the contribution of the BCG is expected to be insignificant. In the WISE 12~$\mu$m and 22~$\mu$m bands the contribution from the central lens galaxy may be significant, and the poor spatial resolution of WISE does not separate the BCG from the lensed source. We thus proceed by using an elliptical galaxy SED model, anchored at the reddest wavelengths at which we have sufficient resolution to measure the central galaxy light independently ($HK$ and 3.6, 4.5~$\mu$m).  
We take as the fiducial SED the Spitzer data on BCGs in clusters from \cite{egami06}. We use the data on central galaxies from the galaxy clusters Abell 773, Abell 2219, and Abell 2261, all of which are  close to \clustername\ in redshift, and are noted in Egami et al.\ as normal central galaxies with no excess flux at long wavelengths.  From these three galaxies, and their K$_s$ magnitudes in the 2MASS catalog \citep{2mass}, we compute colors for the 4.5~$\mu$m, 8~$\mu$m, and 24~$\mu$m bands referenced to the K$_s$ and 3.6~$\mu$m bands. Uncertainties are taken as the spread in these values across the three galaxies. We measure the K$_s$ and 3.6~$\mu$m AB magnitudes of the best fitting BCG model from our data, and then use the colors derived from the data of Egami et al., with simple interpolation, to predict the 12~$\mu$m and 22~$\mu$m magnitudes of the central galaxy in \clustername. Final uncertainties are derived from the uncertainty in the color, added in quadrature to the difference between the predicted magnitudes in the K$_s$ and 3.6~$\mu$m bands. The result is that at 12~$\mu$m the WISE photometry of the lensed source is nominally contaminated at 7\%$\pm$4\% by the central galaxy, and at 22~$\mu$m at 2.5\%$\pm$1.5\%.  In both cases the uncertainty on this correction is less than the photon noise in the measurements.

\subsection{Optical spectroscopy} \label{sec:opspec}
The BCG was observed spectroscopically using the ARC 3.5m telescope and the Double Imaging Spectrograph (DIS) on the night of 8 April 2011, to measure its velocity dispersion.  The 1$\farcs$5 slit was used, with the B1200 and R1200 gratings used on the blue and red channels respectively. The gratings were centered at 4750~\AA\  and 6200~\AA, which provided coverage from 3870~\AA\--5130~\AA\ and 6250~\AA\--7390~\AA\ and nominal resolutions of $R \sim$2000 and $\sim$3000 at the wavelength centers of the blue and red channels. Three 1800~s integrations were acquired under good conditions. Data were reduced to  flux- and wavelength-calibrated stacked 1D spectra using custom IDL scripts incorperating procedures from the XIDL\footnote{http://www.ucolick.org/$\sim$xavier/IDL/index.html} software package. The resulting spectra have per pixel SNRs of typically $\sim10$ and $\sim20$ in the blue and red channels respectively.

\subsection{Near-infrared spectroscopy}  \label{sec:nirspec}
We obtained a spectrum of \arcname\ with the NIRSPEC instrument on the Keck II telescope on the night of 19 Mar 2011 UT.  The weather was clear, with the seeing variable during the night and especially poor while \arcname\ was observed.  The airmass was secz$<$1.07.  The slit position angle was 346 degrees E of N, such that spectra for the two brightest images of \arcname\ were simultaneously captured, as shown in Figure 5. Some light from the BCG is included in the slit though it is spatially distinct from the light from \arcname.  We used the NIRSPEC-2 filter, low resolution mode, a grating angle of 36.48, the 0.76\arcsec\ x 42\arcsec\ slit, and an ABBA nod pattern with a 14\arcsec\ nod.  This setup covered the wavelength range of 1.11 to 1.29~\micron.  We integrated for 600~s at each nod position.  For telluric flux calibration, the star A0~V star HD 131951 was observed immediately after the observations, at an airmass of secz$=$1.06 and a distance of $4^{\circ}$ from \arcname.

We reduced the NIRSPEC spectra using George Becker's pipeline as described in \cite{rigby11}, with one exception: since the emission lines were too faint to extract a one-dimensional spectrum from each nod, we instead summed the sky-subtracted two-dimensional frames taken in Nod A, repeated the summation for Nod B, and then separately extracted the spectra from each summed nod.  For the North image we weighted-average combined the extracted spectra for each of the two nods.  For the South image we used only the spectrum from Nod B, as the spectrum from Nod A suffered persistance from the trace of a telluric star.  The effective integration time was thus 80 min for the North image, and 40 min for the South image.

\section{Results} \label{sec:results}
\subsection{Velocity dispersion of the BCG} \label{sec:vbcg}

The BCG velocity dispersion ($\sigma_{v}$) was obtained using the method of Kelson et al. (2000; see eq.\ 9 therein).  Here, redshift and $\sigma_{v}$ are jointly determined by minimizing $\chi^{2}$ as the galactic spectrum is compared to a shifted, broadened, continuum-matched stellar template in pixel-space using a gradient search algorithm (e.g., \texttt{IDL MPFIT}).  As the SNR of the blue channel BCG data was low, we excluded it from this process, using it only to provide initial redshift information based on Ca H \&\ K. Using both the blue and red channel spectra in their entirety, we find a redshift for the BCG of $z_{BCG} = 0.2373 \pm 0.0002$. Unfortunately, the loss of the blue channel data combined with the BCG redshift and the fixed disperser available to remote observers on DIS removed any overlap between BCG and observed stellar standards, rendering the latter unusable as templates.  
We employed instead a set of high-resolution spectra from the UVES Paranal Observatory Project \citep{bag03}, selected by spectral type without regard for metallicity. 

Before comparison with the BCG spectrum, it was necessary to match the template resolution with that of DIS.  The latter quantity was determined in two ways: 1) gaussian fitting of sky emission lines on 2-D DIS spectra; 2) gaussian broadening of UVES templates and comparison to DIS stellar spectra. The methods yielded values of $\mathrm{FWHM}_{inst} = 1.46 \pm 0.15$\ \AA\ and $1.78 \pm 0.21$\ \AA, respectively, showing no significant trend with wavelength.  We adopted the average of these, $\mathrm{FWHM}_{inst} = 1.62$ \AA\, as the instrumental resolution for DIS, corresponding to $\sigma_{inst}\sim30\, \mathrm{km\ s^{-1}}$ at the relevant wavelengths, observed frame.  All templates were broadened to this resolution before beginning the BCG fitting procedure.

At the SNR of our BCG spectrum, the uncertainties associated with template variation/mismatching (systematic) and photon statistics (random) are dominant and of comparable order. Precisely quantifying the systematic effects requires extensive simulation and is necessary when investigating detailed physical relations based on velocity dispersions, such as
$\mathrm{M}_{b}-\sigma$.  Our interest in this quantity, however, extends only to providing a prior for the lensing model below.  Hence, we deem it sufficient to verify that our procedure adequately reproduces accepted values of $\sigma_{v}$ from the literature. We took as a test sample a set of spectra for 20 Virgo ellipticals from \citet{Dressler84} for which $\sigma_{v}$ measurements were made using a region surrounding the $\lambda5176$ Mg \textit{b} triplet - which we also capture.
The \cite{Dressler84} data are of comparable SNR to our own. Restricting the fitting to the 4900 - 5400 \AA\ interval used in the original paper, the template that best-reduced $\chi^{2}$ across all the \citep{Dressler84} galaxies yielded $-0.02 < \mathrm{log_{10}}(\sigma_{v} / \sigma_{v, D84}) < 0.05$ for $\sigma_{v, D84} \geq 85\ \mathrm{km\ s^{-1}}$ with a difference in the mean consistent with zero. Template-to-template variations introduce a logarithmic scatter of less than 0.1 dex across all $\sigma_{v}$, but do not bias the results. 

Adopting as much of the \cite{Dressler84} interval as our data allow ($\lambda \in [5024,5400]$ \AA) and masking telluric features, we find $\sigma_{v,\ BCG} = 298\ \pm 8\ \mathrm{(random)}\ \pm\ 18\ \mathrm{(systematic)}\ \mathrm{km\ s^{-1}}$. Values are taken from the best-fitting template with the systematic error reflecting RMS variation across templates.

\subsection{Lensing model and magnification} \label{sec:lensmodel}

We construct a lens model of the cluster using the publicly-available software  \texttt{LENSTOOL} (Jullo et al. 2007). The cluster and the BCG are each modeled by a pseudo-isothermal ellipsoid mass distribution (PIEMD; Limousin et al. 2005)\footnote{This profile is formally the same as a dual Pseudo Isothermal Elliptical Mass Distribution (dPIE, see El\'iasd\'ottir et al. 2007).}, parameterized by its position $x$, $y$; a fiducial velocity dispersion $\sigma_{\mathrm{PIEMD}}$; a core radius $r_{core}$; a cut radius $r_{cut}$; ellipticity $e=(a^2−-b^2)/(a^2+b^2)$, where $a$ and $b$ are the semi-major and semi-minor axes, respectively, and a position angle $\theta$. 
We fix the BCG parameters $x$, $y$, $e$, and $\theta$ at their observed values as derived from the GALFIT modeling described above, and fix the cluster cut radius at 1000 kpc as this cannot be constrained by the data. This does not significantly affect the lensing model. 
The BCG velocity dispersion of $\sigma_{v} = 298 \pm 26$~\kms\ measured in \S\ref{sec:vbcg} is set as a Gaussian prior. 
The velocity dispersion of the cluster was not measured, but we can estimate it. \clustername\ appears in the GMBCG catalog (Hao et al. 2010) with a weighted richness $N_{gals}^{\rm weighted}=9.734$; following the calibration of Becker et al. (2007), we estimate a velocity dispersion of $\sigma=318\pm111$~\kms. The low richness and velocity dispersion indicate that \clustername\ is a poor cluster or group, in agreement with the general appearance of the cluster in the optical imaging, and the small separation between the lensed images of the background galaxy. We use this velocity dispersion as a Gaussian prior. Note that the above velocity dispersions are converted to $\sigma_{\mathrm{PIEMD}}$ using the relations in \citep{eliasdottir07}. 
We also note that the BCG in this cluster is very dominant, based on the magnitude difference of 1.64 mag in the $i$-band between the BCG and next two brightest cluster members (for typical poor clusters, this difference is $< 1.0$ mag; \citealt{ned10}). We therefore limit the cluster center to be not farther than 39\arcsec~ from the BCG, which is the separation between the BCG and the next brightest galaxy. Dominant systems tend to show good alignment between the BCG and the cluster halo \citep{ned10}, similar to the results from the lens modeling shown below. 
The best-fit model is found via Monte Carlo Markov Chain (MCMC) optimization in the image plane, using one to four coordinates in each instance of the lensed galaxy as constraints (see Figure 6). The best-fit model has an image-plane RMS of $0\farcs1$. Table~2 lists the best-fit parameters and their $1\sigma$ uncertainties. 

Since the lensed images are extended, the magnification is not fixed across different portions of each lensed image of the source. The average magnification of each lensed image is thus measured as the ratio of areas occupied by the galaxy in the image plane and source plane. In Table~3 we report the average magnification for each image of the source. We also report the magnification of the core regions of the North and South images,
as these regions were targeted for spectroscopy (see Figure 5). Table 3 also includes the total magnification of the aggregate of all images of the source; note that the total magnification of the source is not a linear sum of the magnifications of its instances because the source galaxy crosses the source-plane caustics and some of its lensed instances (the east and west images) represent only parts of the source galaxy rather than all of it. In Figure 7 we show the magnification contours of the best-fit model and a source-plane reconstruction of the galaxy. To estimate the uncertainty on each magnification, we draw parameters from chains in the MCMC process, compute a lens model for each set of parameters, and measure the magnification of each image. The uncertainties quoted in Table 3 are given as the range of magnifications corresponding to $1\sigma$ in parameter space.

\subsection{Redshift, line fluxes, and metallicity} \label{sec:nirspec_analysis}
We separately fit the Keck NIRSPEC spectra for each of the two images of the lensed galaxy, as follows.  The redshift was set initially by fitting a single Gaussian to H-$\alpha$.  We then simultaneously fit the [N~II] 6549.85, 6585.28~\AA\ doublet and the 6564.61~\AA\ H$\alpha$ line \footnote{Wavelengths are vacuum, from NIST at http://www.pa.uky.edu/\~peter/atomic/}, using three Gaussians using the IDL Levenberg--Marquardt least-squares fitting code MPFITFUN \citep{mark09}.  Wavelengths were set at the NIST values\footnote{http://www/pa.uky.edu/$\sim$peter/atomic} and allowed to vary within $\pm2\sigma$ of the initial fit to H-$\alpha$.  A common linewidth for all lines was allowed to vary freely.  The [N~II] doublet flux ratio was fixed at the value computed in \cite{stor2000}.

Table 4 reports the redshift for \arcname\ as measured from simultaneously fitting the [N~II] and H-$\alpha$ lines in each near-IR spectrum.  The result is consistent with that found by fitting H-$\alpha$ alone.  The weighted mean redshift for the two images is $z=0.81589 \pm 0.00005$. The two images have identical redshifts within uncertainties. The spectra are plotted, with their multicomponent fits, in Figure 8. Table 4 lists measured fluxes for H-$\alpha$ and [N~II]. 

We infer metallicities from the measured H-$\alpha$ to [N~II] line flux ratios using the calibration of \cite{pet04}.  The metallicity is calculated for each image of the lensed galaxy, and tabulated in Table 4.  The summed [N~II] and H-$\alpha$ fluxes over both images yields the most precise metallicity: $12 + \log(O/H) = 8.62 \pm 0.06$ using the linear Pettini \& Pagel calibration, and $12 + \log(O/H) = 8.64 \pm 0.09$ using their third-order polynomial calibration.  For reference, solar metallicity is $12+\log(O/H)=8.69 \pm 0.05$ \citep{asp09}.  Thus, we conclude that \arcname\ has solar metallicity.

\subsection{Spectral Energy Distribution Fits} \label{sec:sedfit} 
We fit the rest-frame UV, optical, and near-infrared SED of \arcname\ following the procedure outlined in \citep{eva2010,eva2012}. We use \textit{Hyperz} \citep{bolzonella2000} to perform SED fitting at fixed spectroscopic redshift with the newest Bruzual \& Charlot population synthesis models (CB07 - see Bruzual \& Charlot (2003)), using the Chabrier initial mass function \citep{chab03} and the Calzetti dust extinction law \citep{calz00}. We allow a range of exponentially declining star formation histories with \textit{e}-folding times $\tau$ ranging from 10~Myr to 2~Gyr as well as a constant star formation history. The age of the stellar population is allowed to vary between 50~Myr and the age of the Universe at $z=0.816$.\footnote{6.556 Gyr in our assumed cosmology.}  Informed by the near-IR spectroscopy, we fix the metallicity at the solar value. Figure 9 plots the rest-frame UV to near-IR photometry reported in Table 1 as well as the best-fit stellar population model. From this SED, the derived stellar mass, corrected for lensing, is $\log(\mathrm{M}_*/\mathrm{M}_\odot)=10.8^{+0.3}_{-0.2}$.

The Herschel photometry shows that \arcname\ is extremely bright in the far-IR.  Its infrared spectral energy distribution peaks in $\log{(f_\nu)}$ between 90 and 140 microns in the rest frame, which is a considerably redder peak wavelength than nearby template galaxies like M~82 or Arp~220 (which peak in f$_{\nu}$ at 85~\micron\ and 70~\micron, respectively.)  We fit the 12--500~\micron\ photometry with the local galaxy template set of \cite{riek09}, by convolving each template with the WISE, PACS and SPIRE filter curves.  As shown in Figure 10, the best--fitting template is the $\log{(\mathrm{L(TIR)})}=10.75$~\Lsun\ Rieke template.  (L(TIR) is defined as the rest-frame 8--100~\micron\ luminosity.)   Correcting for lensing magnification \footnote{Note that the two most magnified images in this case (North and South) are complete, and so we do not expect significant differential magnification effects \citep{hev12}.}  and integrating this best-fitting template from rest-frame 8--1000~\micron\ yields an intrinsic far-infrared luminosity of L(TIR)$=4.0^{+1.8}_{-1.1} \times 10^{11}$~\Lsun.  This best-fitting template peaks in f$_{\nu}$ at a rest-frame wavelength of 130~\micron.

It is important to note that the best--fitting template was made from the photometry of galaxies almost a decade less infrared--luminous than is \arcname:    the template's luminosity is $7\times$ lower than our measured L(TIR).  In other words, templates of comparable luminosity provide poor fits; their SEDs are much hotter, peaking in f$_{\nu}$  at rest-wavelengths of 75 to 80~\micron.  Thus, matched-luminosity templates provide a poor fit to the shape of the far-IR SED.  By contrast, a less luminous template with an f$_{\nu}$ peak of 130~\micron, when scaled up in luminosity, provides a much better fit to the far-IR SED.  We will return to this point in \S\ref{sec:discussion}.

\subsection{Star Formation or AGN?} \label{sec:agn}
An important question when deriving physical conditions for \arcname\ is whether there is any significant AGN contribution to the observed emission, either from an AGN in the background lensed galaxy itself, or from an AGN in the central BCG (the foreground lens galaxy).  Here we review the (lack of) evidence for an AGN.   The bulk of the light at near-IR wavelengths and blueward clearly comes from an extended source, both for the lensed source itself and the BCG.  The DIS optical spectrum of the BCG does not show any emission lines: the observed spectra includes the expected wavelength of [OII]~3727. In the NIR spectrum of the lensed source, the H-$\alpha$ line is not unusually broad or strong. Thus, there is no evidence for a significant AGN component in the rest-frame optical or near-IR.  

The reasonable match between the FIR SED and the star-forming template set of \cite{riek09} also gives some encouragement that any putative AGN contribution is no larger than might be disguised in that template set.  The far-IR SED is matched by a composite galaxy template; an AGN would be expected to make hotter dust and thus shift the peak to the blue, which is not observed. We note that the Herschel photometry at 70~$\mu$m (\S\ref{sec:herschel}), which partially resolves the multiple lensed images, also provides a limit on any AGN component from the BCG.  The measured flux in the South image of \arcname\ is $22 \pm 5\%$ of the total flux.  The predicted fraction from the lens model is $31^{+20}_{-12}\%$, and from the measured flux ratios at shorter wavelengths that resolve all components of the source it is $28 \pm 2\%$. Taken at face value, the 70~$\mu$m  and shorter wavelength flux ratios limit the contribution of the BCG at 70~$\mu$m to less than 21\% of the total flux, and are consistent with no contribution at all. Finally, there is also no evidence for emission at the position of \arcname\ in either the FIRST radio data \citep{beck95} or at X-ray wavelengths in the ROSAT All Sky Survey, although these data do not set any strong limits on a low-luminosity AGN contribution. Absent data at {\it any} wavelength that suggest the presence of an AGN, we proceed under the assumption the emission observed from the lensed source is powered by stars only.

\subsection{Star formation rate} \label{sec:sfr}

We estimate the star formation rate (SFR) of \arcname\ from two methods:  H-$\alpha$ and the far-infrared.

To convert the H-$\alpha$ fluxes reported in Table 4 into a star formation rate (SFR) captured by the NIRSpec slit, we apply the calibration of \cite{ken98},  modified for a \cite{chab03} initial mass function:


\begin{equation} SFR = 0.66~L(H\alpha) \times 7.9\times 10^{-42}\end{equation}

With $L(H\alpha)$ specified in ergs~s$^{-1}$ this is the apparent star formation rate in M$_\odot$~yr$^{-1}$ in the slit; it requires an aperture correction, magnification, and extinction correction to get the intrinsic star formation rate.  Poor and variable seeing during the Keck observation means that slit losses are high and uncertain.  We estimate the seeing by convolving the J-band image of the brightest cluster galaxy until it matches the width of the BCG in the two-D spectrum, and then compute the slit losses by examining the seeing-degraded images of the source. We conclude that the aperture correction is large:  a factor of $4 \pm 1$. Final intrinsic SFRs from the H-$\alpha$ measurements are computed for each image combining uncertainties in the line flux (Table 4), the magnification (Table 3), and the aperture correction. We find SFR = $31^{+21}_{-11}$ M$_\odot$~yr$^{-1}$ from the North image and SFR = $25^{+16}_{-9}$ M$_\odot$~yr$^{-1}$ from the South image. Note these are both measurements of the total SFR for the entire source; combining with equal weights yields a final best value of SFR$_{H-\alpha}$ = $28^{+13}_{-7}$ M$_\odot$~yr$^{-1}$.



We also estimate the star formation rate from the far-infrared luminosity, using the template fitting discussed in \S\ref{sec:sedfit} and plotted in Figure~10.  The integrated rest-frame 8 to 1000~\micron\ luminosity, corrected by the total magnification, is 4.0$^{+1.8}_{-1.1} \times 10^{11}$ \Lsun.  Via the calibration of \cite{ken98} this corresponds to a SFR of 41$^{+18}_{-11}$ M$_{\odot}$~yr$^{-1}$.

\subsection{Extinction}
If we attribute to extinction the higher value of the SFR measured from the FIR versus H-$\alpha$, then the extinction at H-$\alpha$ is $0.4^{+0.7}_{-0.4}$~mag.  The extinction at H-$\alpha$ estimated from the SED fit is $0.9^{+0.7}_{-0.4}$~mag.  Thus, \arcname\ may have slightly higher extinction than the calibrating sample of \citet{ken98}, though only at $1\sigma$ significance.

\subsection{Physical Size and Structural Properties} \label{sec:size}
The good image quality of the ground-based imaging of \arcname\ allows a robust measure of the galaxy's size and morphology. The strong lensing model for \arcname\ also shows unusually low distortion for the images of this source, compared, for example, to the highly magnified giant arcs more typical in strong lensing galaxy clusters \citep[e.g.,][]{ker12}; this suggests that measurements in the image plane can be simply converted to the source plane. To thoroughly explore the source morphology and size, we have fit the North and South images of the galaxy both directly in the image plane and also in reconstructed images in the source plane. We consider the $JHK_s$ images as well as the $i$-band image; these are the deepest and best seeing images, spanning two distinct instruments and telescopes. Visual examination of the initial images of \arcname\ (see Figure 3) suggests a two-arm spiral morphology for the source. In fitting the source using simple models we mask the obvious portions of these apparent spiral arms.

To fit in the image plane we use GALFIT and the PSF references already established for modeling the BCG light, and fit the isolated (and complete) North and South images of \arcname\ independently in the BCG-subtracted frames. Fits were made using both a single S\'{e}rsic profile,\footnote{An exponential disk has a S\'{e}rsic index of 1, and a de Vaucouleur profile has a S\'{e}rsic index of 4.} as well as a single Gaussian profile to facilitate comparison to the measurements of \cite{ruj11}. Length scales are converted to the source plane by dividing by the square root of the magnification; in the case of equal magnification radially and tangentially, as is approximately true here, this is appropriate. The results of this fitting are reported in Table 5. Treating each image in each filter with equal weight, the measured S\'{e}rsic index in the image plane is 0.87$\pm$0.14 with an effective radius of 4.1$\pm$0.6~kpc and a Gaussian size (in $\sigma$) of 7.7$\pm$0.8~kpc. To measure morphology in the source plane we construct source plane images from the image plane using the lensing model, by simple interpolation on the irregular grid of image pixels produced in the source plane. This is not optimal from a signal to noise perspective since it treats all image plane pixels equally, regardless of magnification; it is however sufficient for the purposes of this paper. To obtain a PSF reference, we map the image plane PSFs back to the source plane in the same way, at the position of the center of the source. Formally, in the source, the realized PSF is a function of position; again, because the distortions are not strong, this effect is sub-dominant. We have tested a range of PSFs sampled from locations around the center of the source, and find no significant effect on the results. Again treating each image in each filter with equal weight, the measured S\'{e}rsic index in the source plane is 1.23$\pm$0.25 with an effective radius of 4.3$\pm$0.6~kpc and a Gaussian size of 7.1$\pm$0.9~kpc.

These measurements, which are consistent across four filters, two images of the source, and both the image plane and the source plane, clearly show that \arcname\ is dominated by a disk with an exponential light distribution.  This is consistent with the visual impression of the system as a two arm ``grand design" spiral galaxy.  The nucleus of \arcname\ also appears significantly elongated, suggesting a bar.

\subsection{Color and color gradients} \label{sec:color}

Figures 6 and 7 include both color composite images as well as greyscale images reconstructed in the source plane in various filters. 
The images show that \arcname\ has an overall red and uniform morphology. The most significant deviations from color uniformity occurs in one of the putative spiral arms, which shows a few bluer clumps. Note that only the core of \arcname was targeted for near-IR spectroscopy and shown to have vigorous star formation - the uniform red light seen across much of the source is that of a vigorously star forming disk. A possible alternate interpretation of the source is as a merger with both arms representing tidal tails;  blue tidal features with many young stars are seen in mergers nearby \citep[e.g.,][]{whit10}. However, the good fit to a exponential disk model, and the otherwise uniform colors of this source disfavor the merger hypothesis. Higher resolution imagery will be required to further illuminate this question.


\section{Discussion and Conclusions}\label{sec:discussion}

The far-IR luminosity of \arcname\ makes it a ``Luminous Infrared Galaxy'' or LIRG.   It is instructive to place \arcname\ in the context of the deep Spitzer surveys done using the MIPS 24~\micron\ band.  The majority of galaxies detected in those surveys have redshifts of  $0.6<z<1.2$ \citep{lefloch05}.  At z$=$0.8 the typical infrared luminosity for MIPS detections is L(TIR)$\sim1\times10^{11}$ \Lsun, with uncertainties of order of a factor of 2-3 \citep{lefloch05}.  The redshift and luminosity of \arcname\ of $z=0.816$ and L(TIR)$= 4.0\times10^{11}$~L$_\odot$   mean that were \arcname\ not lensed, it would appear as a rather typical actively star--forming galaxy in the Spitzer deep fields.  Such galaxies dominate the star formation rate density of the universe at that epoch \citep{papovich04,lefloch05}

\citet{rujo-seds} succinctly summarize results from the past decade showing that actively star--forming galaxies in the distant universe are different from $z=0$ analogues of comparable luminosity in at least three major ways.  First, their infrared SEDs are colder.    Second, their aromatic features grow stronger with redshift for fixed L(TIR).  Third, their star formation occurs on physically larger scales.  

The first two effects were discovered first, and a common cause was suspected: that if star formation was happening on large spatial scales, that would naturally explain the higher aromatic feature strengths (less extinction) and colder SEDs (less intense radiation fields).   New measurements of large sizes for $z>1$ LIRGs and ULIRGs \citep{ruj11} confirm those predictions.  Thus, we have a framework to understand how  the population of vigorously star--forming galaxies evolve with redshift.  Though \arcname\ is only one galaxy, its high lensing magnification enables us to test this framework at otherwise unavailable photometric quality and spatial resolution.  Thus, we now summarize the evidence for evolution with redshift in the SED shape, aromatic strength, and size of star-forming regions in luminous infrared galaxies, and discuss how \arcname\  tests this picture.

{\bf Cooler SEDs:}  That luminous infrared galaxies in the distant universe may have colder SEDs\footnote{As is typical in the U/LIRG literature, we use the words ``hotter'' and ``cooler'' SEDs as a shorthand, meaning that the peak of the far-IR spectral energy distribution is shifted to shorter wavelengths or longer wavelengths, respectively.} than at $z=0$ was suggested by \citet{efs03} and first measured by \citet{rowan-robinson04} from ISO data for LIRGs at $0.15<z<0.5$, and then confirmed with Spitzer \citep{rowan-robinson05,sajina06,symeonidis}.  Spitzer 70 and 160~\micron\ photometry showed that LIRGs and ULIRGs at $z\sim1$ have, on average, colder far-infrared SEDs that peak at longer wavelengths than $z\sim0$ analogues \citep{symeonidis}.  Figure 5 of \citet{symeonidis} illustrates how $0.1<z<1$ LIRGs have longer--wavelength SED peaks  (in $f_{\nu}$) than $z=0$ LIRGs.  The SED peak of \arcname\ is entirely consistent with their distant LIRG sample, and inconsistent with the $z=0$ sample.  Those authors note that a simple way to describe the evolution in the luminosity-SED relation with redshift is that  ``the far-IR SEDs of many high-redshift sources may resemble more closely those of lower-luminosity galaxies locally.''  This result was better quantified with multiband photometry from Herschel:  \citet{elbaz10} and \citet{hwang10} found that the median dust temperature of LIRGs at $z\sim1$ is several K colder than $z=0$ analogues; the offset is even more pronounced at higher luminosities.   

The dust-reprocessed SED of \arcname\ is entirely consistent with these trends.  We fit the SED with the local template set of \citet{riek09}, with the luminosity scaling as a free parameter.  The best--fitting template belongs to $z=0$ galaxies that are nearly an order of magnitude less IR-luminous than \arcname; that template peaks in f$_{\nu}$ at 130~\micron.  By contrast, the Rieke templates which bracket the observed L(TIR) luminosity of \arcname\ have f$_{\nu}$ peaks at 75 to 80~\micron.  Thus, the dust SED of \arcname\ looks like a cool, low--luminosity galaxy that has been ``scaled up'' in luminosity.

{\bf Aromatic feature strength:}  The evolution toward stronger aromatic features with redshift, at fixed L(TIR), was first discovered from Spitzer mid--infrared spectra and far-IR spectral energy distributions of $1<z<3$ galaxies \citep{papovich07,rigby08,farrah08,karin09}.  The explanation proposed by the latter three sets of authors was that star formation in distant galaxies may occur on larger spatial scales than in local galaxies.  Unfortunately, for \arcname\ we cannot isolate the aromatic feature strengths, as we lack mid-infrared spectra for this source.  The WISE photometry at 12 and 22~\micron\ do contain flux from the aromatic features, but blended in unknown ratio with hot dust continuum.  Thus, it is not possible to test whether the aromatic features in \arcname\ obey the evolutionary relation noted in \citet{rigby08}.  

{\bf Size:}  \citet{ruj11} measured Gaussian sizes of $0.4<z<2.5$ LIRGs and ULIRGs of 2--10~kpc.  These are considerably larger than the the measured $\la 1$~kpc sizes of star formation in comparable--luminosity galaxies at $z=0$.  Thus, \citet{ruj11} conclude that luminous star formation episodes at $z=0$ are found only on sub-galactic ($D<1$~kpc) spatial scales, while at $z>0.4$ they occur on larger, galaxy--wide scales ($D\sim3$--8~kpc), with consequently lower star formation rate surface densities.  

The Gaussian size of \arcname\ (7.4$\pm$0.6~kpc) is slightly larger than the sizes of 3--6~kpc measured by \citet{ruj11} for $z\sim1$ LIRGs, but consistent given our uncertainties and their small sample size.   We can also compare to size measurements from HST imaging of disk galaxies in the GEMS fields. \cite{bar05} predict a relationship between effective radius (of a S\'{e}rsic profile fit) and total stellar mass.  For the measured stellar mass of \arcname\ of $\sim 6\times10^{10}$ \Msun, the predicted effective radius from their mean relation is $\sim7$~kpc.   This is somewhat larger than the measured value of 4.3$\pm$0.4~kpc for \arcname, though consistent with expectations given the significant spread in the values of effective radius at fixed stellar mass apparent in the results of \cite{bar05}. 

{\bf\arcname\ in this context:}  
\arcname\ is only a single galaxy, but the strong lensing magnification allows for a detailed suite of measurements otherwise impossible to obtain.  The redshift, luminosity, and far-IR SED of \arcname\ show it to be a typical vigorously star forming galaxy at $z\sim0.8$, and thus representative of those galaxies that contribute most of the universe's star formation at this earlier epoch.  [Its dust--reprocessed SED, which we measure at high signal-to-noise-ratio, is clearly cooler than comparably luminous galaxies at $z=0$, as are its peers at similar redshift].  The spatial extent of star formation in \arcname\ is much larger than in comparably luminous galaxies at $z=0$, and comparable to (or slightly larger than) the half-dozen LIRGs measured at $0.5<z<1$ by \citet{ruj11}, confirming that result at higher signal-to-noise and spatial resolution. The observations of \arcname\ presented here knit together both the size and SED-to-luminosity evolution seen piecemeal in other work in a single, and apparently typical, object.

What is unique about \arcname\ is that lensing magnification allows us to determine the morphology of this vigorous star formation at high spatial resolution and signal-to-noise ratio. Morphological measurements, both in the image plane and the reconstructed source plane, strongly suggest that the star formation in \arcname\ is occurring in a large exponential disk. The surprising result, as show in Figure 7, is that star formation appears to be distributed reasonably smoothly over the entire disk of this galaxy. 

Thus, in this single but typical object, by robustly measuring a size, a dust temperature, and a total infrared luminosity, we confirm the picture from previous work that star formation in luminous infrared galaxies at z$=$1 occurs on much larger physical scales than in comparable-luminosity galaxies at z$=$0.    Further, at high spatial resolution in \arcname, we see that star formation is taking place smoothly across an exponential disk with an effective radius of 4.3$\pm$0.6~kpc and an appearance suggestive of a two-armed 'grand design' spiral morphology.  The only somewhat comparable object in the local volume is HIZOA~J0836-43 \citep{culver08,culver10} -- a LIRG that similarly shows a large disk, a cool FIR SED, and a SFR of $\sim$21 M$_\odot$~yr$^{-1}$; HIZOA~J0836-43 is extremely gas rich (M$_{HI}$ = 7.5$\times$10$^{10}$ M$_\odot$ : \citealt{culver08}) suggesting that \arcname\ may be an excellent target for future observations of molecular gas content, for example with ALMA or the EVLA.

\appendix

\section{The red herring}
The discovery of cluster \clustername\ as a strong lens has a curious history, in that the initially--identified arc candidate proved not to be lensed at all, while the follow-up infrared imagery revealed a different, very red galaxy that was also suggested in the SDSS imaging, is actually lensed, and is the subject of this paper.  For completeness, we now describe the initial arc candidate.

The lensing search process detailed in Gladders et al.\ (in prep.) does not record which feature caused a particular field to be flagged as a candidate for lensing; the most likely feature is this arclet-like source, though a second faint and red feature - the lensed source discussed in this paper - is just visible in the SDSS imaging. A portion of the custom SDSS color image used in the lensing search of Gladders et al. is reproduced in Figure 1.

Follow-up imaging from the Nordic Optical Telescope (Figure 1) strengthened the lensing interpretation for the candidate arclet, since the curvature suggested in the SDSS imaging is confirmed, and the image does not show any bulge-like component which might be expected in the case of an edge-on galaxy.

The candidate blue arclet was also observed spectroscopically using the ARC 3.5m telescope and the Double Imaging Spectrograph (DIS) on March 2nd, 2008. These data show that the candidate blue arclet is {\it not} in fact lensed; rather, it has a redshift of $z=0.231$ from clear detections of [OII]~3727 and H-$\alpha$,  which is consistent with this source being a cluster member, given the SDSS redshift for a nearby apparent cluster galaxy at $z=0.237$ as well as the redshift of the BCG of 0.2373 (see \S4.1).

\acknowledgments
LA thanks Alan Dressler for extensive discussions regarding velocity dispersion measurements, and for kindly providing spectra on Virgo early-type galaxies. We thank the Herschel Science Centre for Director's Discretionary time to observe \arcname.

MDG thanks the Research Corporation for support of this work through a Cottrell Scholars award.

This publication makes use of data products from the Wide-field Infrared Survey Explorer, which is a joint project of the University of California, Los Angeles, and the Jet Propulsion Laboratory/California Institute of Technology, funded by the National Aeronautics and Space Administration (NASA).

This work made use of observations made with the Spitzer Space Telescope, which is operated by the Jet Propulsion Laboratory, California Institute of Technology under a contract with NASA. Partial support for this work was provided by NASA through an award issued by JPL/Caltech.

This work includes observations obtained at the Gemini Observatory, which is operated by the 
Association of Universities for Research in Astronomy, Inc., under a cooperative agreement 
with the National Science Foundation (NSF) on behalf of the Gemini partnership: the NSF (United 
States), the Science and Technology Facilities Council (United Kingdom), the 
National Research Council (Canada), CONICYT (Chile), the Australian Research Council (Australia), 
Ministério da Ciência, Tecnologia e Inovação (Brazil) 
and Ministerio de Ciencia, Tecnología e Innovación Productiva (Argentina).

This work was partially supported by a NASA Keck PI Data Award, administered by the NASA Exoplanet Science Institute.  This work includes data obtained at the W. M. Keck Observatory from telescope time allocated to NASA through the agency's scientific partnership with the California Institute of Technology and the University of California. The Observatory was made possible by the generous financial support of the W. M. Keck Foundation.

This publication makes use of data products from the Two Micron All Sky Survey, which is a joint project of the University of Massachusetts and the Infrared Processing and Analysis Center/California Institute of Technology, funded by NASA and the NSF. 

We acknowledge the use of data from the UVES Paranal Observatory Project (ESO DDT Program ID 266.D-5655).

This paper makes use of the ROSAT Data Archive of the Max-Planck-Institut f\"{u}r extraterrestrische Physik (MPE) at Garching, Germany.

The authors wish to acknowledge the very significant cultural role and reverence that the summit of Mauna Kea has always had within the indigenous Hawaiian community. We are most fortunate to have the opportunity to conduct observations from this mountain.

\clearpage

\begin{deluxetable}{lllllllll}
\tabletypesize{\scriptsize}
\tablecolumns{6}
\tablewidth{0pc}
\tablenum{1}
\tablecaption{Imaging observations and photometry of \arcname\label{tab:obs}}
\tablehead{
\colhead{Telescope \& Instrument} & \colhead{filter} & \colhead{t$_{int}$ (s)} &
\colhead{PSF (\arcsec)} & \colhead{date} & \colhead{m$_{AB}$}}
\startdata
Gemini-N  GMOS                 & SDSS g & 300  & 0.61        & 29 Apr 2011 & 21.45$\pm$0.16 \\ 
Gemini-N  GMOS                 & SDSS r & 300  & 0.56        & 29 Apr 2011 & 20.24$\pm$0.11 \\
Gemini-N  GMOS                 & SDSS i & 300  & 0.54        & 29 Apr 2011 & 18.85$\pm$0.10 \\
Gemini-N  GMOS                 & SDSS z & 300  & 0.50        & 29 Apr 2011 & 18.10$\pm$0.09 \\
Magellan  FourStar             & Y      & 800  & 0.68        & 19 Feb 2011 & 18.03$\pm$0.12 \\
Magellan  FourStar             & J      & 800  & 0.66        & 19 Feb 2011 & 17.34$\pm$0.11 \\
Magellan  FourStar             & H      & 800  & 0.63        & 19 Feb 2011 & 16.84$\pm$0.11 \\
Magellan  FourStar             & K$_s$  & 1572 & 0.53        & 23 Jan 2011 & 16.31$\pm$0.11\\
Spitzer IRAC\tablenotemark{a}  & 3.6~\micron & 600 & 1.7     & 19 Sep 2010 & 15.74$\pm$0.13  \\
Spitzer IRAC\tablenotemark{a}  & 4.5~\micron & 600 & 1.7     & 19 Sep 2010 & 16.14$\pm$0.13  \\
WISE                           & 12~\micron  & 167 &  6.5& Jan/Jul 2010   & 14.58$\pm$0.08\tablenotemark{c}\\
WISE                           & 22~\micron  & 193 & 12.0& Jan/Jul 2010   & 14.16$\pm$0.21\tablenotemark{c}\\
Herschel PACS\tablenotemark{b} & 70~\micron  & 900 & 5.6     & 09 Jul 2011 & 12.06$\pm$0.15 \\
Herschel PACS\tablenotemark{b} & 100~\micron & 900 & 6.8     & 09 Jul 2011 & 11.11$\pm$0.15 \\
Herschel PACS\tablenotemark{b} & 160~\micron &1800 & 11      & 09 Jul 2011 & 10.00$\pm$0.15 \\
Herschel SPIRE\tablenotemark{b}& 250~\micron & 185 & 18      & 12 Jul 2011 & 10.06$\pm$0.15\\
Herschel SPIRE\tablenotemark{b}& 350~\micron & 185 & 25      & 12 Jul 2011 & 10.41$\pm$0.15\\
Herschel SPIRE\tablenotemark{b}& 500~\micron & 185 & 36      & 12 Jul 2011 & 11.34$\pm$0.15 \\
\enddata
\tablecomments{Columns are: 
1) Telescope and Instrument; 
2) filter or bandpass; 
3) integration time in seconds; 
4) point spread function, quoted as the seeing disk full width at half maximum (FWHM) for ground-based instruments, 
and the tabulated point spread function FWHM for space observatories;
5) universal date of observation;
6) AB magnitude for all images of the source in aggregate - see main text for further details;}
\tablenotetext{a}{Spitzer data are from Program 70154, ``Mass across the Redshift Desert: 
Stellar Masses in a Large and Uniform Sample of Strongly-Lensed Galaxies at $1<z<3$'' (PI Gladders).}  
\tablenotetext{b}{Herschel images are from program DDT\_jrigby\_2, ``The far-IR SED of a highly magnified red galaxy''
(PI Rigby).}
\tablenotetext{c}{Central galaxy flux subtracted using SED model, as detailed in main text.}

\end{deluxetable}


\begin{deluxetable*}{lccccccc} 
\tabletypesize{\footnotesize}
 \tablecolumns{8} 
\tablenum{2}
\tablecaption{Best-fit lens model parameters  \label{tab:lmod}} 
\tablehead{\colhead{Halo }   & 
            \colhead{RA}     & 
            \colhead{Dec}    & 
            \colhead{$e$}    & 
            \colhead{$\theta$}       & 
            \colhead{$r_{\rm core}$} &  
            \colhead{$r_{\rm cut}$}  &  
            \colhead{$\sigma_{PIEMD}$}\\ 
            \colhead{(PIEMD)}   & 
            \colhead{($\arcsec$)}     & 
            \colhead{($\arcsec$)}     & 
            \colhead{}    & 
            \colhead{(deg)}       & 
            \colhead{(kpc)} &  
            \colhead{(kpc)}  &  
            \colhead{(km s$^{-1}$)}  } 
\startdata 
Cluster  & \Px   & \Py   & \Pe   & \Ptheta    & \Prc   & [1000] & \Psigma  \\ 
BCG       & [0] & [0] & [0.64]  & [167]  & \Prcb & \Pcutb     & \Psigmab  \\ 
\enddata 
 \tablecomments{All coordinates are measured in arcseconds relative to the center of the BCG, at [RA, Dec]=[219.68768 14.903507]. The ellipticity is expressed as $e=(a^2-b^2)/(a^2+b^2)$. $\theta$ is measured north of West. Error bars correspond to 1 $\sigma$ confidence level as inferred from the MCMC optimization. Values in square brackets are for parameters that are not optimized. 
}
\end{deluxetable*} 
\begin{deluxetable}{lll}
\tabletypesize{\footnotesize}
\tablecolumns{7}
\tablewidth{0pc}
\tablenum{3}
\tablecaption{Magnifications of images of SGAS~143845.1+145407\label{tab:mags}}
\tablehead{
\colhead {Image} & \colhead{Magnification} & \colhead{Notes}}
\startdata
North       &  \magN & complete image\\
South       &  \magS & complete image\\
East        &  \magE & partial image\\
West        &  \magW & partial image\\
North-Core  &  \magcoreN & NIR slit region\\
South-Core  &  \magcoreS & NIR slit region\\
All         &  \magall & \\
\enddata
\tablecomments{Columns are: 1) a source image name - see Figure~6 2) the estimated magnification from the strong lensing model, and 3) notes about the nature of each image. Due to the lensing configuration two images are complete images of the source, and the other two are partial images only. The magnification of the `All' image gives the magnification of all four images in aggregate relative to a single unlensed instance of the source; this is the appropriate magnification to use when converting the observed photometry in Table~1 back to unlensed quantities.
}
\end{deluxetable}

\begin{deluxetable}{lllllll}
\tabletypesize{\footnotesize}
\tablecolumns{7}
\tablewidth{0pc}
\tablenum{4}
\tablecaption{Measured emission lines in SGAS~143845.1+145407\label{tab:HA_fluxes}}
\tablehead{
\colhead {object} & \colhead{redshift} & \colhead{f(H-$\alpha$)} & \colhead{f(NII 6585)} & \colhead{[N II]/H-$\alpha$} & \colhead{12+log(O/H)} &  \colhead{12+log(O/H)}}
\startdata
North image       &   $0.81588 \pm 0.00006$  &   $229  \pm 16$ &  $83 \pm 14$  &    $0.36  \pm  0.07$  & $8.65^{+0.04}_{-0.05}$ & $8.69^{+0.08}_{-0.09}$\\
South image       &   $0.8159  \pm 0.0001$   &   $238  \pm 28$ &  $69 \pm 20$  &    $0.29  \pm  0.12$  & $8.59^{+0.07}_{-0.09}$ & $8.59^{+0.12}_{-0.14}$\\
both images     &   $0.81589 \pm 0.00005$  &   $467  \pm 32$ &  $152\pm 24$  &    $0.33  \pm  0.07$    & $8.62^{+0.05}_{-0.06}$ & $8.64^{+0.09}_{-0.09}$\\
\enddata
\tablecomments{Columns are: 1) object name, 2) measured redshift and uncertainty from the 
three-component fit to H-$\alpha$ and [N II], 3) H-$\alpha$ flux and uncertainty, 
4) [N II] 6585~\AA\ flux and uncertainty, 5) ratio of [N II]/H~$\alpha$ and uncertainty, 
6) derived oxygen abundance using the linear N2/H-$\alpha$ calibration of Pettini \& Pagel 2004; 
7) derived oxygen abundance using the third-order N2/H-$\alpha$ calibration of Pettini \& Pagel 2004.
Reported fluxes are in units of $10^{-17}$~\cgsflux.  
We separately quote measured values for the North and South images.  Where we quote 
values for ``both images'', these are the 
weighted mean and error in the mean over all components for the redshift, 
the sum over all components for each flux, and the ratio of the summed fluxes for the line ratio.
}
\end{deluxetable}

\begin{deluxetable}{llllllll}
\tabletypesize{\footnotesize}
\tablecolumns{8}
\tablewidth{0pc}
\tablenum{5}
\tablecaption{Morphology and Size of SGAS~143845.1+145407, S\'{e}rsic and Gaussian Profile Fits\label{tab:morph}}
\tablehead{
\multicolumn{2}{c}{}  & \multicolumn{3}{c}{Image Plane} & \multicolumn{3}{c}{Source Plane} \\
\colhead{Image} & \colhead{Filter} & \colhead{R$_e$ (kpc)} & \colhead{S\'{e}rsic Index} & \colhead {Gaussian $\sigma$ (kpc)} & \colhead{R$_e$ (kpc)} & \colhead{S\'{e}rsic Index} &\colhead {Gaussian $\sigma$ (kpc)} }
\startdata

South Image       & K$_s$ & 3.6 & 0.99 & 7.0 & 3.7 & 1.19 & 7.0\\
                  & H     & 3.6 & 0.96 & 7.2 & 4.6 & 1.50 & 8.2\\
                  & J     & 4.0 & 0.71 & 8.4 & 4.8 & 1.66 & 8.4\\
                  & SDSS-i& 4.3 & 0.74 & 7.7 & 5.1 & 1.22 & 7.0\\ \hline
&&&&&&&\\
North Image       & K$_s$ & 4.0 & 1.01 & 6.9 & 3.9 & 1.18 & 6.1\\
                  & H     & 4.1 & 1.01 & 7.3 & 3.8 & 1.20 & 6.2\\
                  & J     & 4.9 & 0.73 & 9.1 & 3.7 & 1.03 & 6.4\\
                  & SDSS-i& 4.6 & 0.78 & 7.8 & 4.5 & 0.87 & 7.2\\ \hline
&&&&&&&\\
Average           & all   & 4.1$\pm$0.6 & 0.87$\pm$0.14 & 7.7$\pm$0.8 & 4.3$\pm$0.6 & 1.23$\pm$0.25 & 7.1$\pm$0.9\\  
\enddata
\tablecomments{Columns are: 1) source image name 2) filter, 3)--5) effective radius, S\'{e}rsic index and Gaussian size in the source plane from fits in the image plane, and 6)--8) effective radius, S\'{e}rsic index and Gaussian size from fits to the reconstructed source plane images. In the case of the image plane, effective radii and Gaussian sizes are computed in the source plane simply by dividing by the square root of the magnification, as this is the limiting case suggested by the only modest distortions seen in the magnified images in the image plane. No such correction is applied to fits in the source plane, however the treatment of the PSF is commensurately more complex, as detailed in the main text.
}
\end{deluxetable}

\clearpage

\begin{figure}
\figurenum{1}
\includegraphics[width=3.2in]{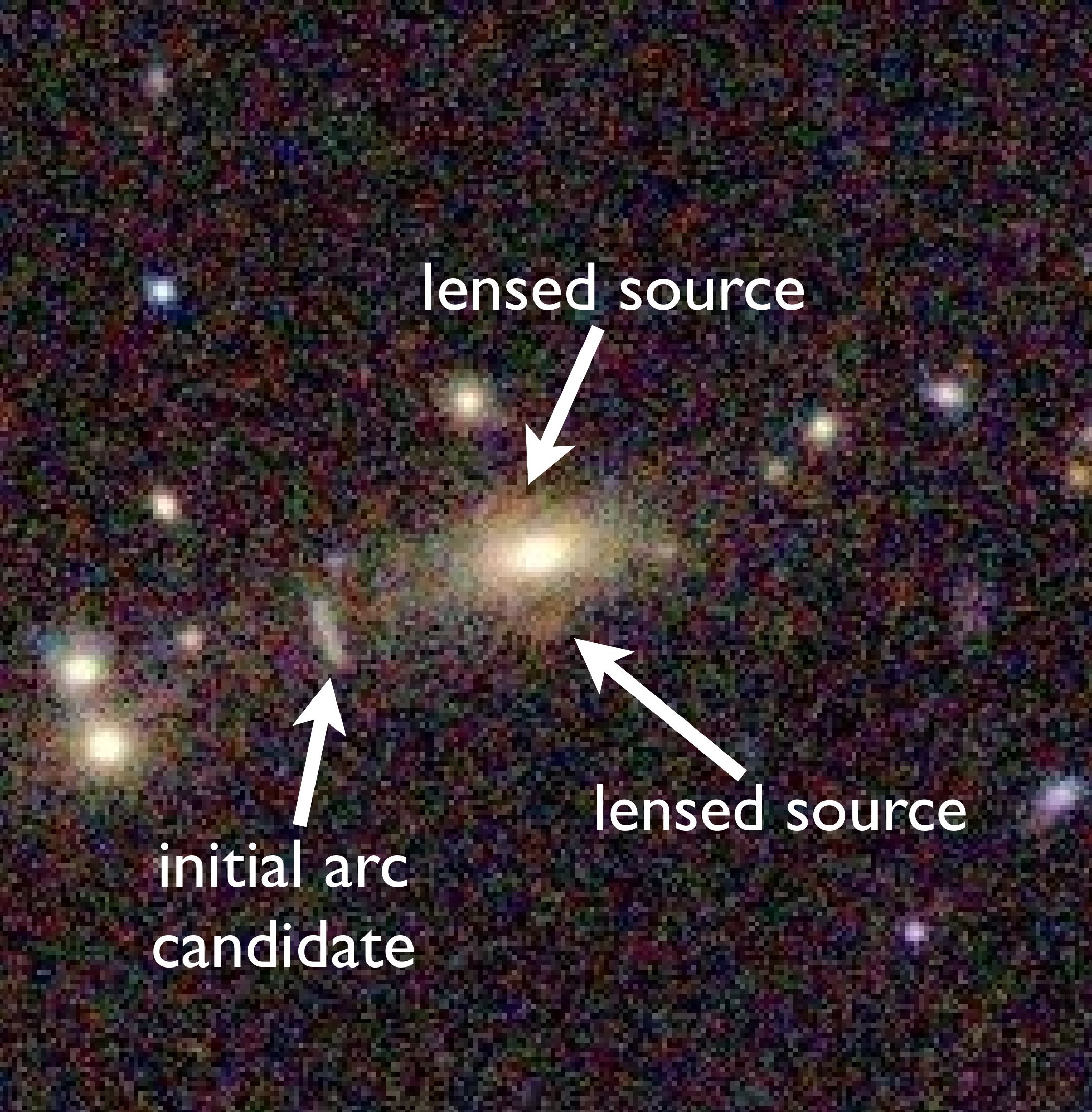}
\figcaption{A 90x90\arcsec\ image cutout from SDSS DR7 imaging data (blue: $g$-band, green: $r$-band, red : $i$-band + $z$-band).  The faint red light around the BCG, which is barely visible all but to the right of the BCG, is the lensed source discussed in this paper.  The arclet to the lower left is the candidate lensed source discussed in Appendix A.}\label{fig:sdss_cutout}
\end{figure}

\begin{figure}
\figurenum{2}
\includegraphics[width=5.0in]{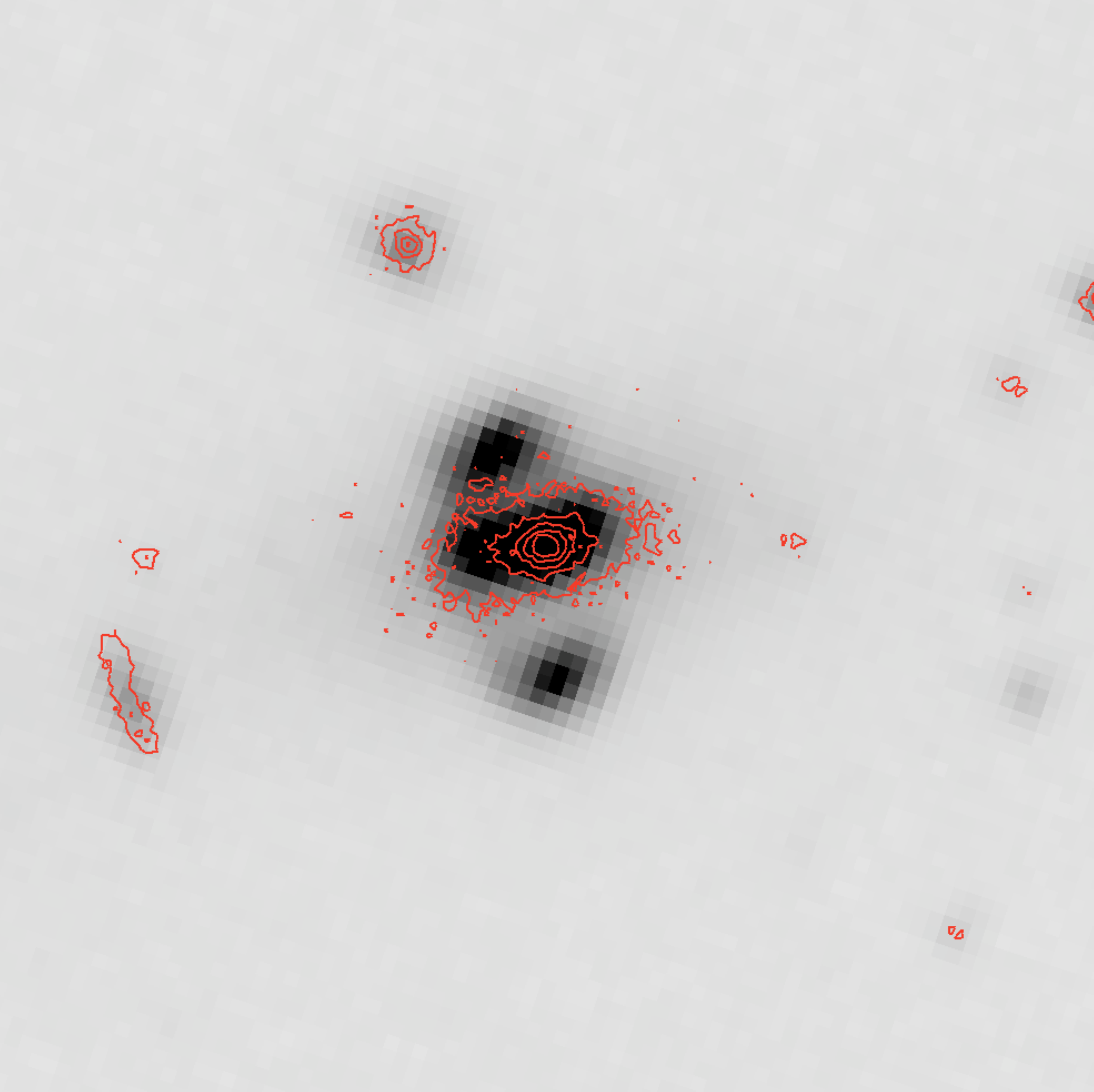}
\figcaption{A 45x45\arcsec\ cutout from the Spitzer IRAC 3.6~$\mu$m image (greyscale), with the NOT g-band image over-plotted (red contours).  Two extremely red sources above and below the BCG are immediately apparent; these are two images of \arcname.}\label{fig:spitzer_cutout}
\end{figure}

\clearpage

\begin{figure}
\figurenum{3}
\includegraphics[width=7.2in]{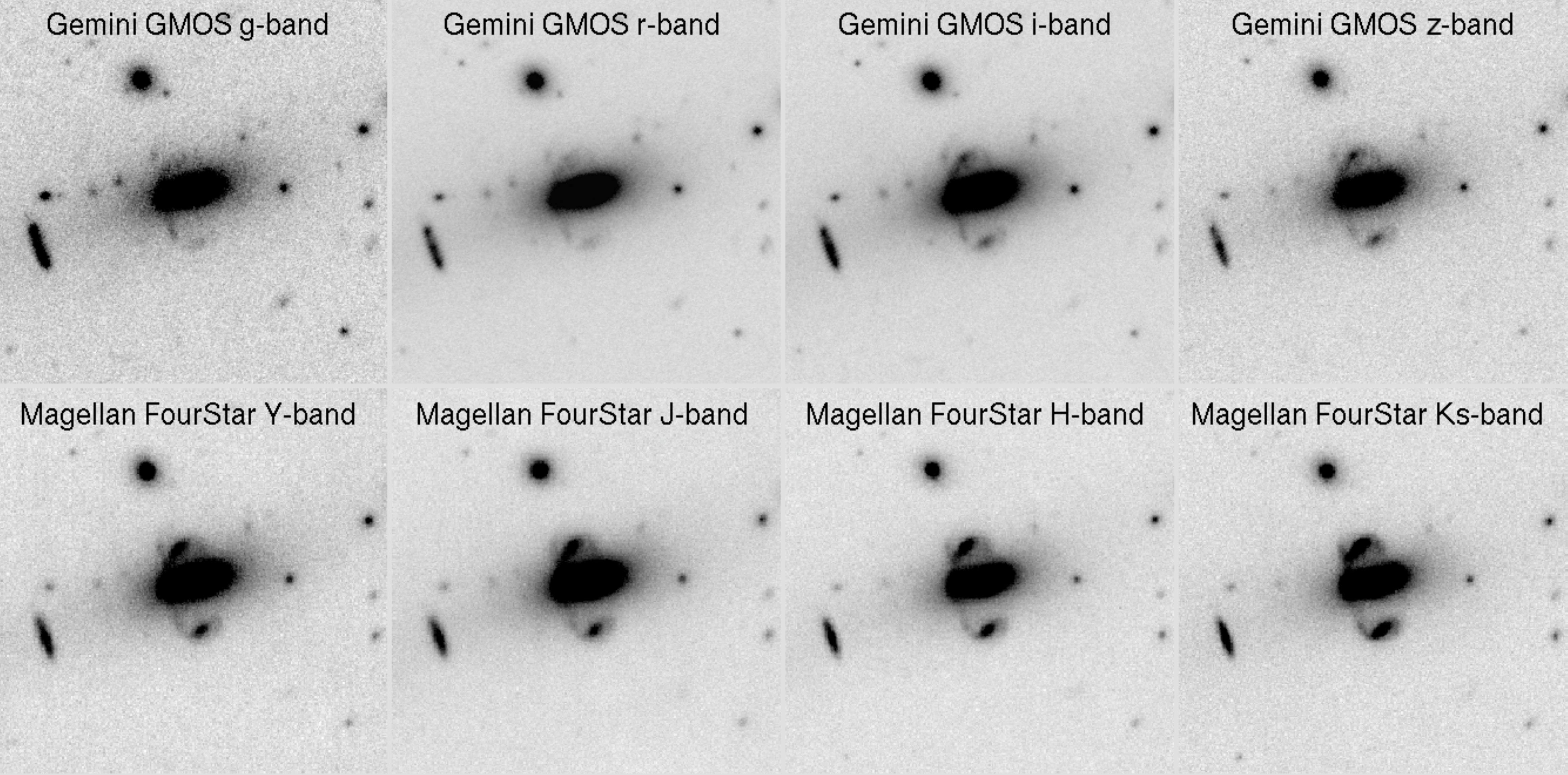}
\figcaption{45x45\arcsec\ cutouts in the $grizYJH$ and $K_s$ bands. All images have been scaled to approximately the same greyscale level for the BCG. Note the obvious multiplicity of images, particularly in the near-IR, which confirms the lensing interpretation.}\label{fig:manybands}
\end{figure}

\begin{figure}
\figurenum{4}
\includegraphics[width=5.0in]{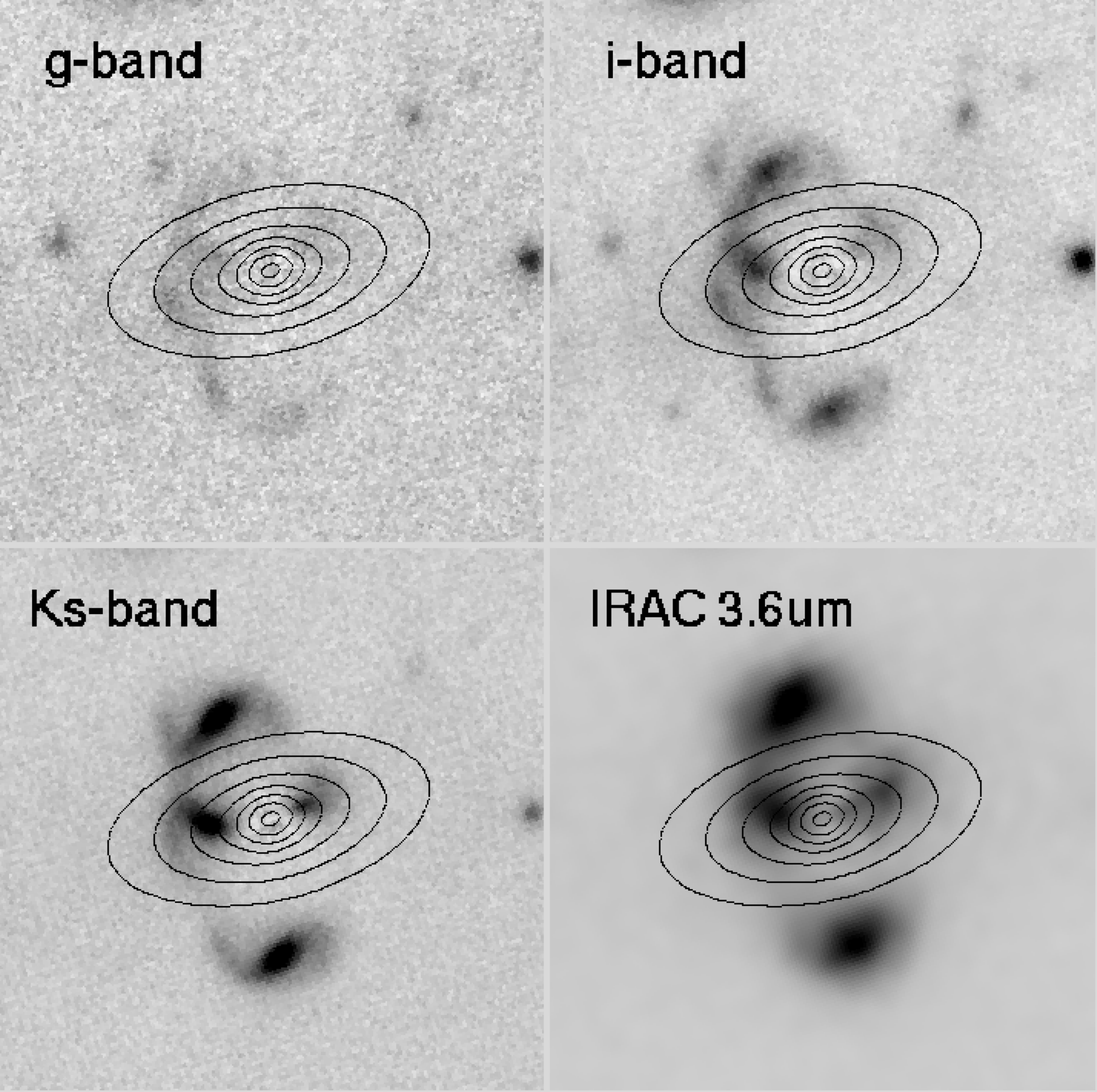}
\figcaption{Example images (20x20\arcsec) after subtraction of the brightest cluster galaxy (BCG).  Residuals near the center of the field are modest, with amplitudes that are $1$--$2\%$ of the BCG core prior to subtraction. The BCG location and orientation is indicated by contours, taken from theK$_s$-band model.}\label{fig:bcgsub}
\end{figure}

\clearpage

\begin{figure}
\figurenum{5}
\includegraphics[width=3in]{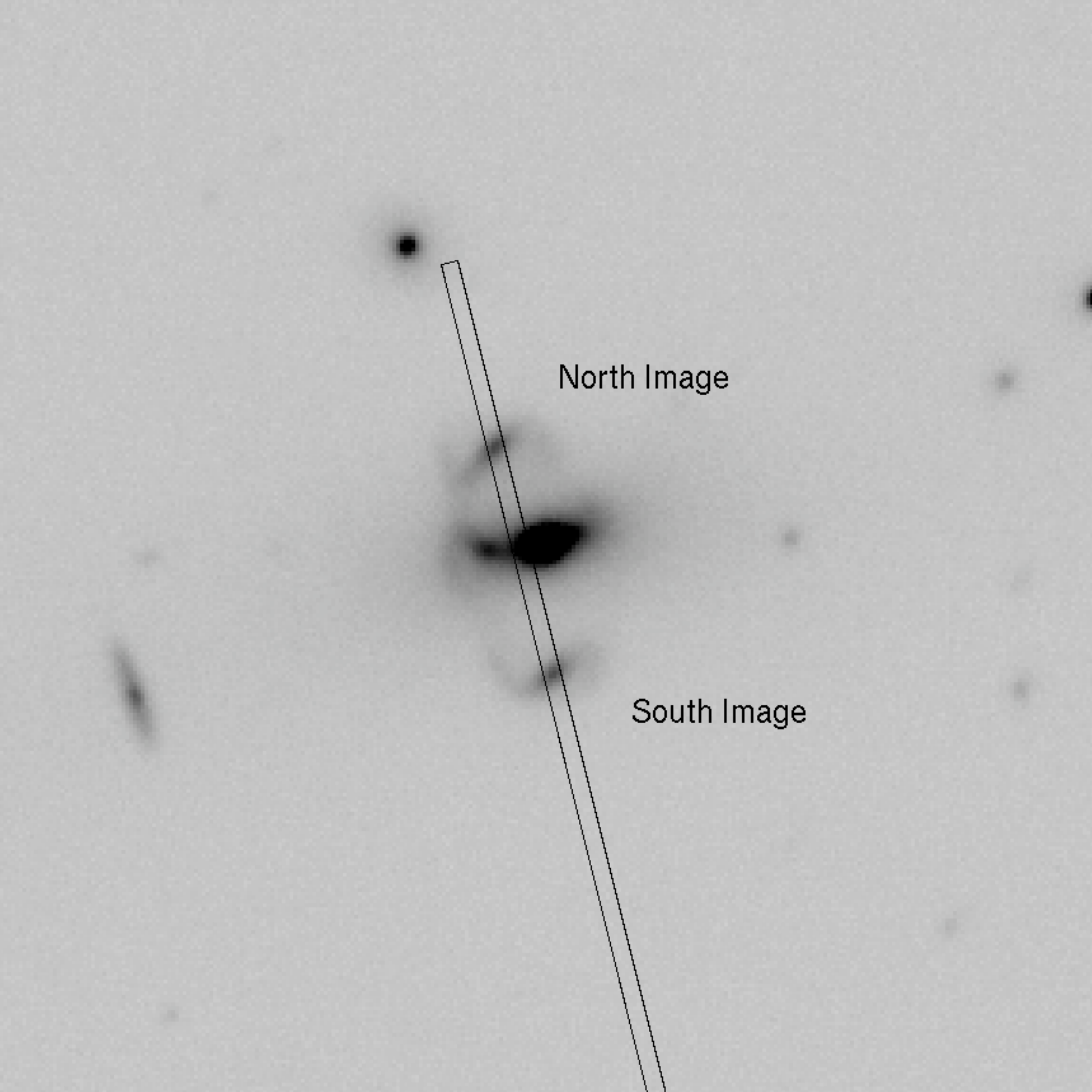}
\figcaption{A 45x45\arcsec\ K$_s$-band image showing how the Keck NIRSPEC slit was placed to cover two images of the lensed source.  The slit is placed at one of the two nod positions.} \label{fig:nirspec_slit}
\end{figure}

\begin{figure}
\figurenum{6}
\includegraphics[width=5in]{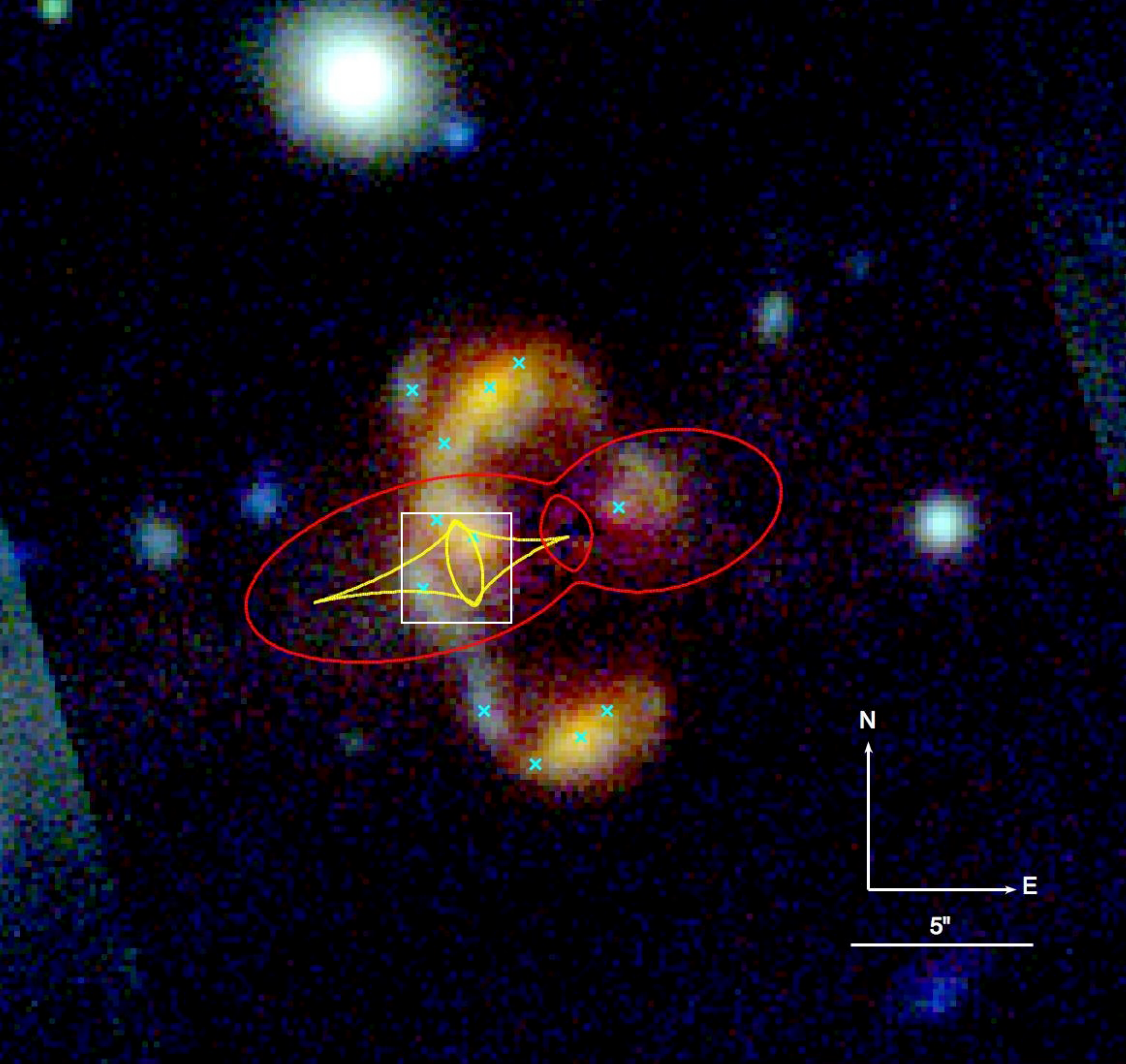}
\figcaption{The lens model: critical curves (red lines) and caustics (yellow lines) from the best-fit lens model are overplotted on a color composite image (red: $J$, green: $i$, blue: $g$). The coordinates within each arc that were used as lensing constraints are marked in cyan. 
The location of the source in the source plane (see Figure~\ref{fig:sourceplane}) is indicated by a white box.
}
\end{figure}

\clearpage

\begin{figure}
\figurenum{7}
\includegraphics[width=3.2in]{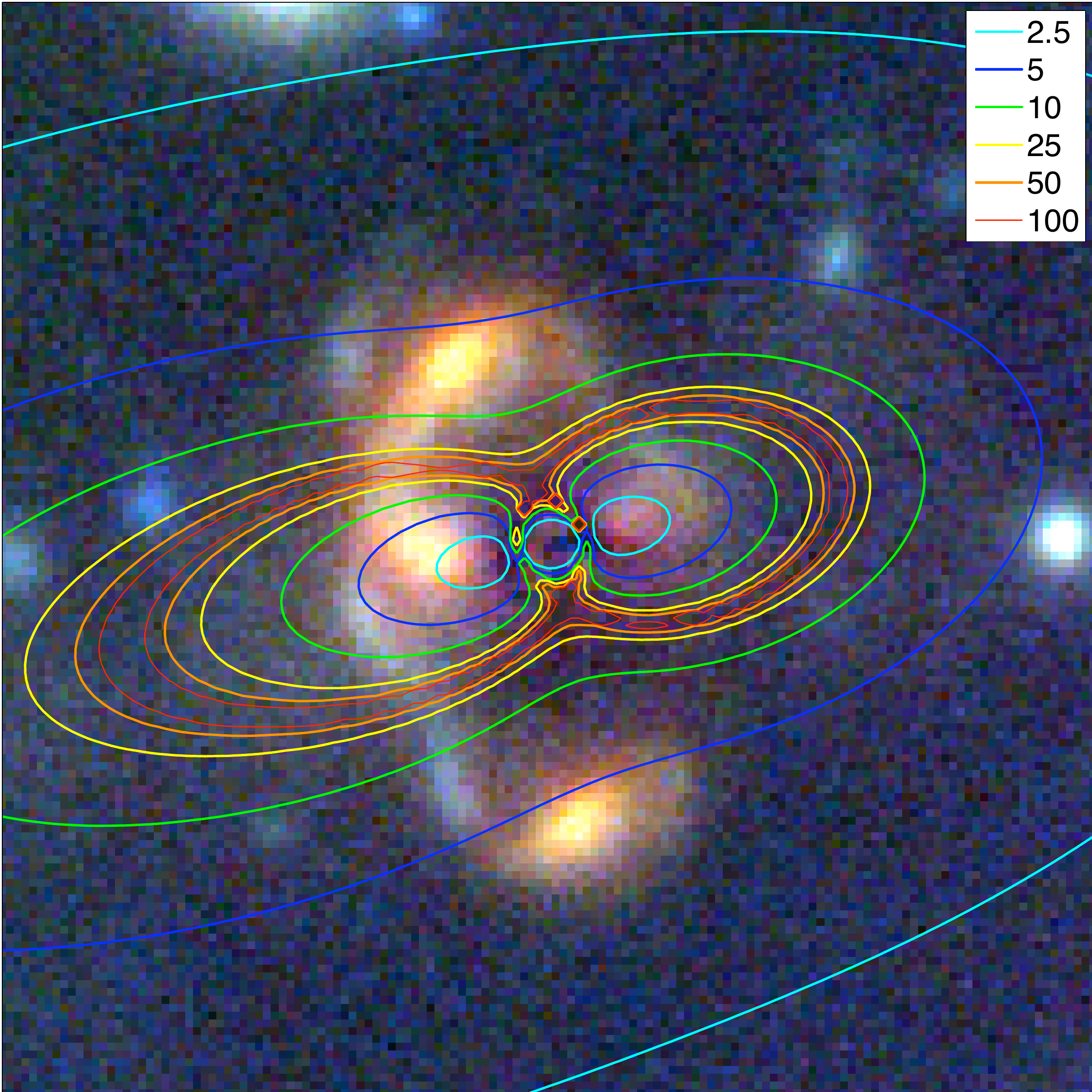}
\includegraphics[width=3.2in]{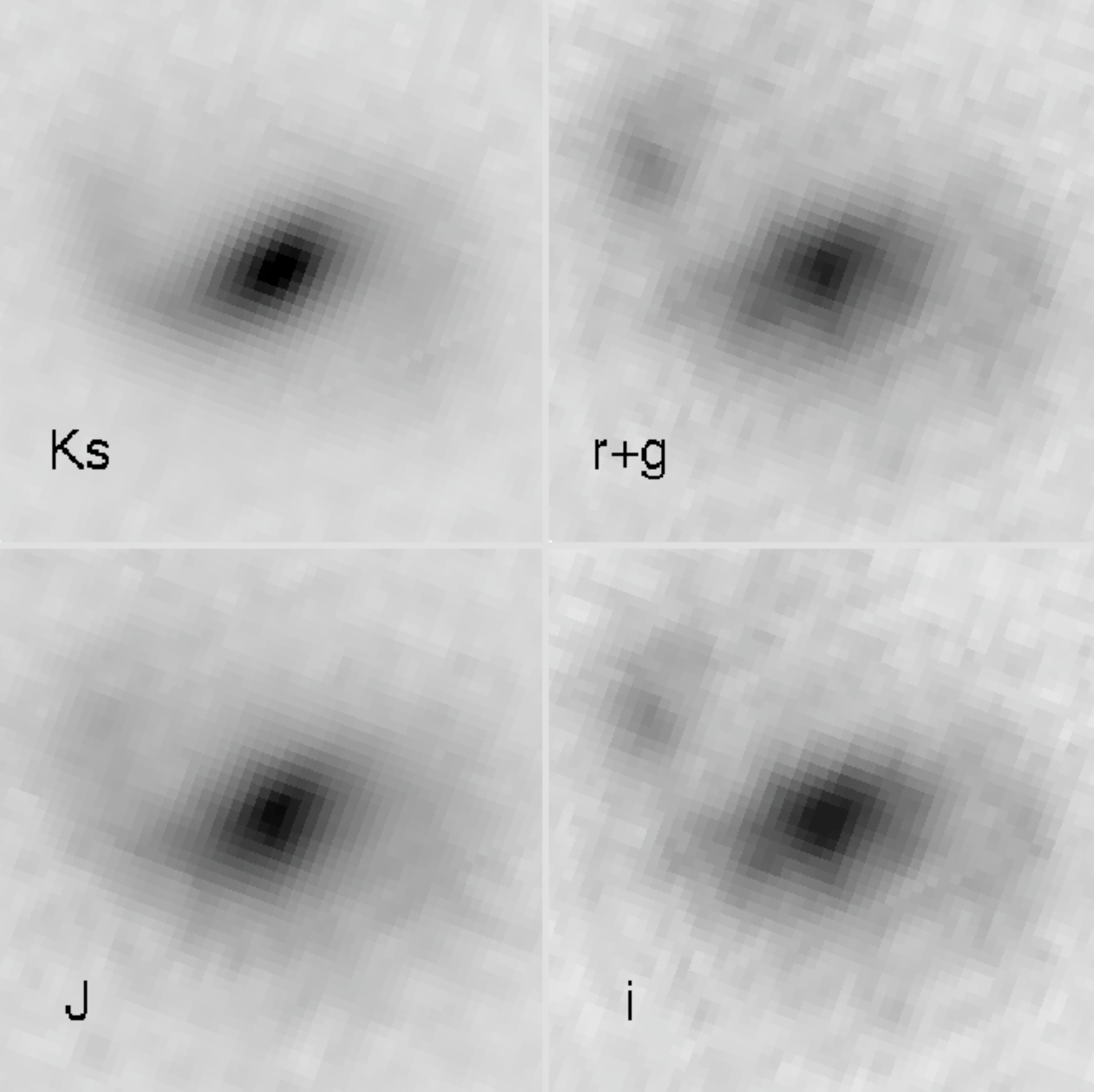}
\figcaption{The lens model: {\it left:} Contours of absolute magnification are overplotted on a color composite image (red: $J$, green: $i$, blue: $g$) of \arcname. 
{\it Right:} Reconstruction of the source of \arcname, from the best-fit lens model. We ray-trace each pixel in the image plane through the lens model, and resample the source plane into a $0\farcs05$ pixel grid.}\label{fig:sourceplane}
\end{figure}

\begin{figure}
\figurenum{8}
\includegraphics[width=3.0in,angle=0]{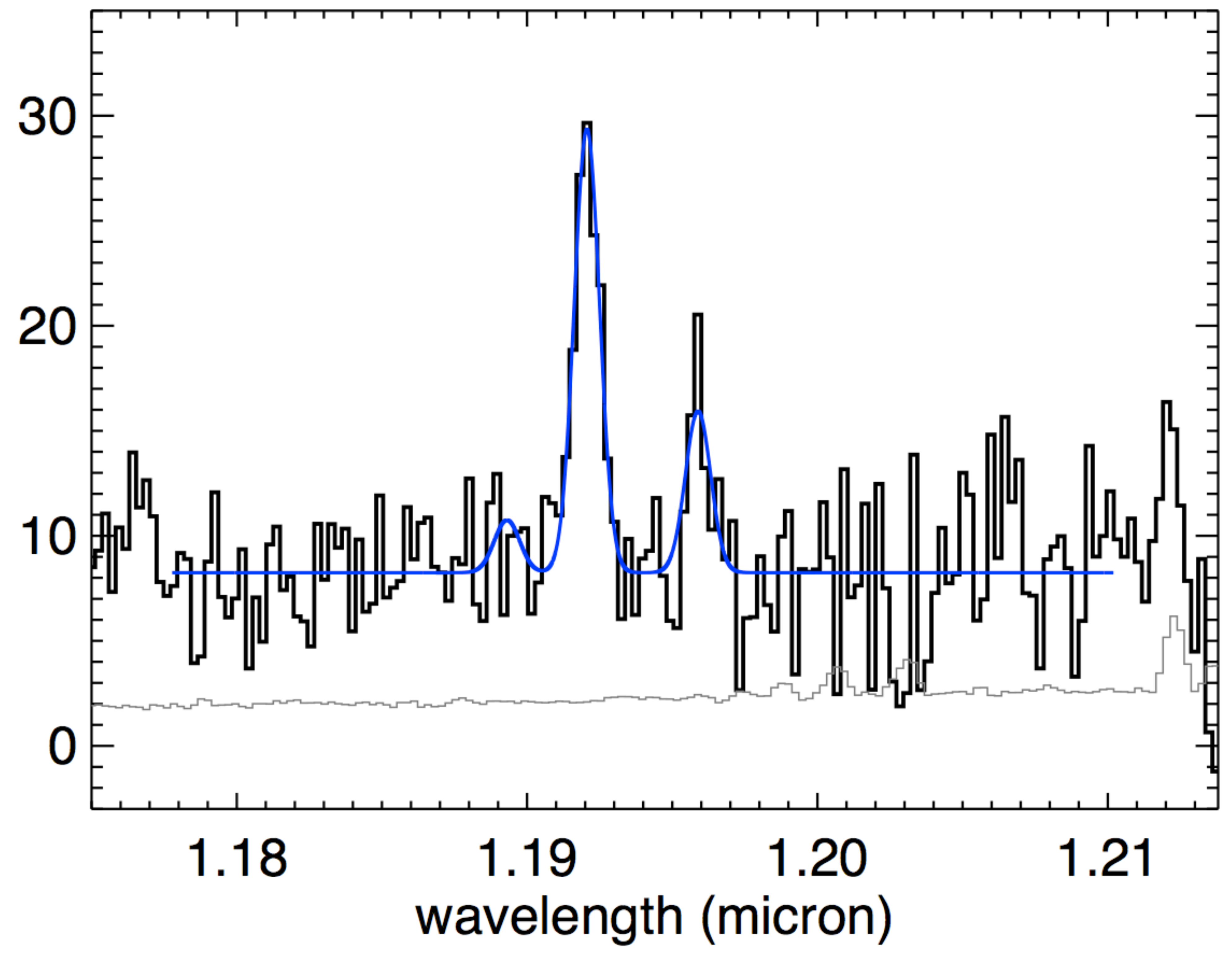}
\includegraphics[width=3.0in,angle=0]{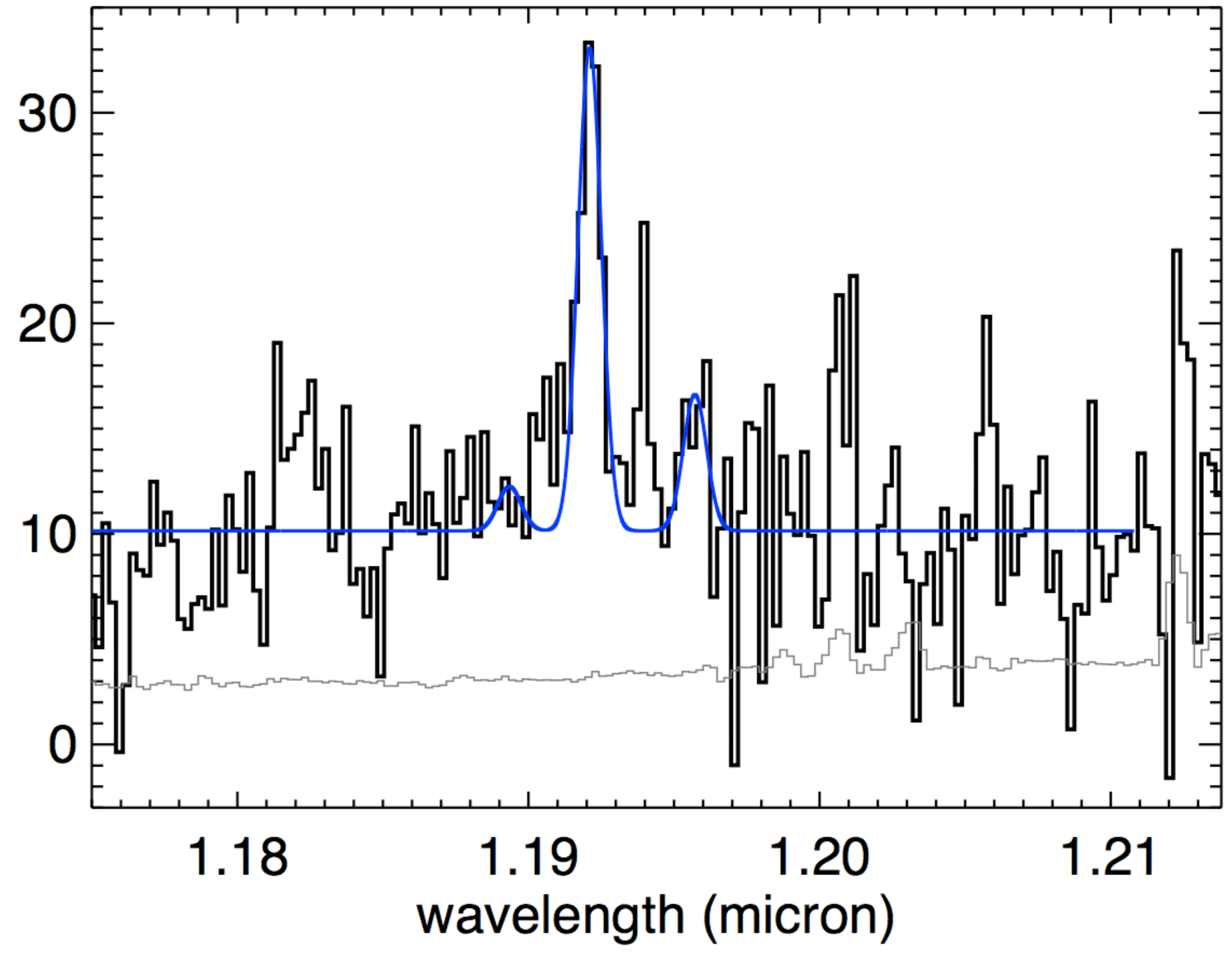}
\figcaption{Observed Keck/NIRSPEC spectra of H$\alpha$ and [N II] in \arcname.  Spectra of the North and South images are plotted in the left and right panels, respectively.  Each X-axis shows observed wavelength in \micron, and each Y-axis shows
observed specific intensity $f_{\lambda}$ in units of $10^{-17}$~\cgsflam.
In each panel, the spectrum is plotted in black, the 1$\sigma$ uncertainty spectrum is plotted in grey, and the multi-component Gaussian fit is plotted in blue.  The extent of the blue line shows the region over which the continuum was fit.  In each spectrum, the three fitted emission lines are [N II] 6548, H$\alpha$, [N II] 6583.}\label{fig:IRspec}
\end{figure}

\begin{figure}
\figurenum{9}
\includegraphics[width=6in]{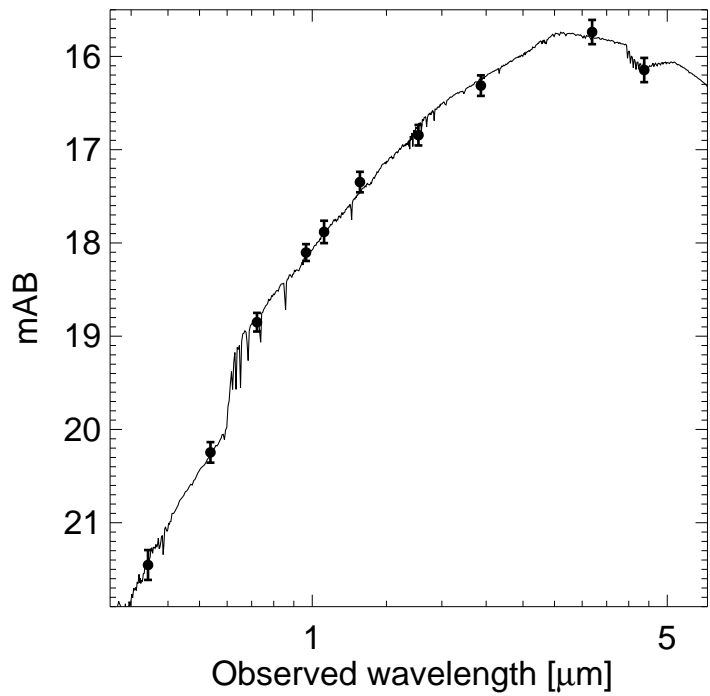}
\figcaption{The photospheric spectral energy distribution of the lensed source \arcname.  Datapoints correspond to $gri$ photometry from Gemini, $YJHK_s$ photometry from Magellan, and 3.6-4.5~$\mu$m photometry from Spitzer, as detailed in Table 1 and described in \S\ref{sec:data}. Also plotted is the best-fitting Bruzual \& Charlot stellar population synthesis model, as described in \S\ref{sec:sedfit}.  The units of the Y axis are AB magnitudes.  Photometry has been summed over all images of \arcname.}
\end{figure}

\begin{figure}
\figurenum{10}
\includegraphics[width=6in]{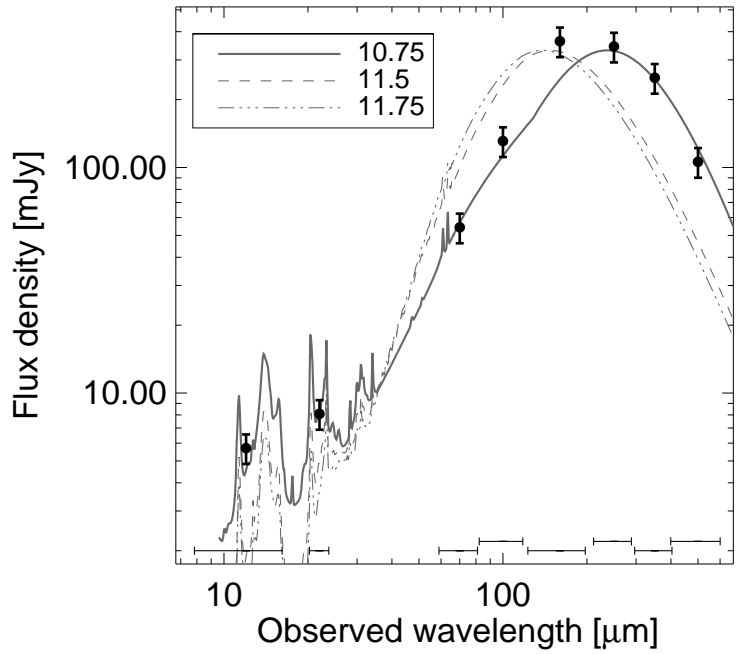}
\figcaption{The dust--reprocessed spectral energy distribution of the lensed source \arcname.  Datapoints correspond to observed photometry at observed wavelengths of 12 and 22~$\mu$m from WISE, at 70, 100, and 160~$\mu$m from Herschel/PACS, and at 250, 350, and 500~$\mu$m from Herschel/SPIRE, as detailed in Table 1. Wavelength ranges for each observation are plotted toward the bottom. Also plotted is the best--fitting local--galaxy template \textit{(solid line)} from \citet{riek09}, which is the $\log{(L(TIR))}=10.75$~\Lsun\ template.  We measure the L(TIR) of \arcname\ as $7 \times$ higher than this fiducial template; in other words, the template that best fits \arcname\ was made from relatively low--luminosity galaxies, and has been scaled up in luminosity by a factor of 7 to fit a galaxy with a much higher luminosity.  To illustrate this, we plot for comparison the Rieke templates whose luminosities bracket that of the source:  $\log{(L(TIR))}=11.5$ and $11.75$ \Lsun; these have been normalized to the same peak flux density as the best-fitting template.  Clearly these templates are hotter, and do not fit the SED shape of \arcname.}
\end{figure}


\end{document}